\documentclass[aps,jcp,reprint,twocolumn,citeautoscript,showpacs,longbibliography]{revtex4}
\usepackage{graphicx}
\usepackage{caption}
\usepackage{subcaption}
\usepackage{amsmath}
\usepackage{amssymb}
\usepackage{tabularx}
\usepackage{placeins}
\usepackage{units}
\usepackage{color}
\usepackage{hyperref}
\usepackage{ulem}

\usepackage{dcolumn}
\usepackage{bm}

\begin{document}

\title{A new generation of effective core potentials for correlated calculations}

\author{M. Chandler Bennett$^{1,2,*}$, Cody A. Melton$^{1,2,*}$, Abdulgani Annaberdiyev$^{1}$, Guangming Wang$^{1}$,
Luke Shulenburger$^{2}$, and Lubos Mitas$^1$}
\affiliation{
1) Department of Physics, North Carolina State University, Raleigh, North Carolina 27695-8202, USA \\
}
\affiliation{
2) Sandia National Laboratories, Albuquerque, New Mexico 87123, USA \\
}

\thanks{These authors contributed equally to this work}

\date{\today}


\begin{abstract}
We outline ideas on desired properties for a new generation of effective core potentials (ECPs) that will allow valence-only calculations to reach the full potential offered by recent advances in many-body wave function methods.
The key improvements include consistent use of correlated methods throughout ECP constructions and improved transferability as required for an accurate description of molecular systems over a range of geometries. 
The guiding principle is the isospectrality 
of all-electron and ECP Hamiltonians for a subset of valence states. We illustrate these concepts on a few first- and second-row atoms (B, C, N, O, S) and we obtain
higher accuracy in transferability than
previous constructions while using a semi-local ECPs with a small number of parameters.
In addition, the constructed ECPs enable many-body calculations of valence properties with higher (or same) accuracy than their all-electron counterparts with uncorrelated cores.  This implies that the ECPs include also some of the impacts of core-core and core-valence correlations
on valence properties. The results open further prospects for ECP improvements and refinements. 
\end{abstract}

\maketitle
In recent decades, many-body electronic structure methods have enjoyed steady progression in accuracy and efficiency; this has largely been driven by algorithmic improvements coupled with the increasing power of computational resources.
These advances can be readily seen, for instance, in methods that rely on wave function expansions, such as coupled cluster (CC) methods,
as well as stochastic approaches that sample either particle configurations or the Fock space of antisymmetric wave functions.
For example, quantum Monte Carlo (QMC) methods that utilize explicitly correlated many-body wave functions often reach accuracies of 1-2~kcal/mol for systems containing elements from several rows of the periodic table \cite{Foulkes:2001rmp,Kolorenc:2011rpp}.
For practical reasons these methods often approximate properties of atomic cores, especially when calculating systems with heavier elements. 
In particular, the cores are either left uncorrelated, fully frozen or even eliminated entirely using pseudopotentials or effective core potentials (ECPs) \cite{foot1}.
Note, however, that ECPs are generated within approximate theories as they are often tuned to match Density Functional Theory (DFT) or Hartree-Fock (HF) all-electron atomic properties in the valence space.

In correlated treatments of very heavy elements, the use of ECPs is essentially unavoidable 
since explicit correlation of atomic cores would become another bottleneck for such calculations. For instance, in QMC the corresponding computational cost grows as $\mathcal{O}(Z^{5-6})$ where $Z$ is the nuclear charge.
Although ECPs provide an important efficiency boost, they inevitably introduce errors that can compromise the overall quality of results. 
Furthermore, the increasing accuracy of the many-body methods mentioned above reveals that ECP imperfections can become a dominating source of systematic bias \cite{Shulenburger:2013prb,nazarov}.

In recent decades a number of ECP tables as well as on-demand computational packages have been established using several types of electronic structure approaches. 
This includes DFT constructions such as the comprehensive table of Bachelet et al \cite{Bachelet} based on relativistic DFT
and ECPs generated by the OPIUM package \cite{OPIUM}. We also mention recent advances in DFT constructions with
multiple projectors \cite{Hamann2013}. Several sets of tables suitable for many-body calculations have been developed using 
Hartree-Fock/Dirac-Fock (HF/DF) methods \cite{STU,DolgCao,SBK,CRENBL,CRENBL2,HW,Burkatzki:2007jcp,Trail:2005jcp}. 

Most of the existing ECP constructions have relied on reproducing one-particle properties such as norm/charge-conservation
\cite{Bachelet} or closely related shape consistency \cite{CRENBL,SBK,LES}, HF/Dirac-Fock (DF) excitation energies\cite{STU}
or their extensions \cite{Hamann2013} including a recent composite scheme designed to accurately build-in many-body effects \cite{Trail:2013jcp,Trail:2015jcp,Trail:2017jcp}. Comprehensive reviews \cite{Pickett,DolgCao}
contain details as well as many relevant references which we skip here. 
To the best of our knowledge, the incorporation
of many-body approaches into ECP constructions have been rather limited
so far; first by the QMC work of Acioli and Ceperley \cite{Acioli:1994jcp}, also the works of Maron, et al.\cite{Maron:cp1998} and Fromager et al.\cite{Fromager:jcp2004} and recent works of Trail and Needs \cite{Trail:2013jcp,Trail:2015jcp,Trail:2017jcp}.
Overall, ECPs have arguably been some of the most successful ways to eliminate unnecessary degrees of freedom through effective Hamiltonians. They have helped to save huge amounts of computational time and effort over the past few decades, and are largely responsible for enabling well-converged many-body solid state calculations.

As indicated above, the current generation of ECPs is becoming a significant hurdle in achieving a level of accuracy consistent with high-accuracy all-electron correlated calculations.
Presently, valence-only correlated approaches often require additional ECP testing, redeveloping, adapting, modifying, and revalidation in order to get the systematic errors originating in ECPs below a desired threshold.
Such practice is highly undesirable and indeed hampers a sizable fraction of QMC and other correlated wave function calculations. This situation is similar to that of past DFT calculations, where pseudopotentials were largely constructed by hand. Optimization approaches \cite{SchilpfCPC2015} and systematic testing\cite{delta_codes_paper,Lejaeghereaad3000} versus all-electron approaches and solid-state calculations have recently been very successful and show that there is a significant room for improvements. However, the substantially higher cost of many-body calculations makes a direct use of these approaches for ECP development quite challenging. 

In this work, we envision and assess practical approaches for constructing a new generation of ECPs.
We have several goals in mind.
First, an effective Hamiltonian that includes an ECP is much more useful if it is reasonably simple and constructed for broad use in basically all theories, i.e., that its universality is similar to the original all-electron Hamiltonian. 
One goal, therefore, is to probe for effective and simple forms that are appropriate for use in both mean-field and/or explicitly correlated methods.
In this way, the effective ECP Hamiltonian provides a well-defined reference so that systematic errors from subsequent approximations and approaches are on full display and can be clearly benchmarked.

Another key goal is to increase the accuracy of the new ECPs beyond previous constructions, using  measures that we define later. 
We explore a few strategies for constructing ECPs using correlated wave function methods from the outset. 
Ideally, these effective Hamiltonians should reproduce, as closely as possible, the behavior of the valence electrons with the original Hamiltonian in a vast range of chemical environments; regardless of whether one calculates a molecule or a condensed system, in an equilibrium or non-equilibrium conformation, and within weak or strong bonding settings.
Therefore, our goal is to construct an effective Hamiltonian that mimics the many-body valence spectrum, the spatial structure of eigenstates and overall scattering properties of the original, relativistic, all-electron atom's Hamiltonian. 
Clearly, some compromises will have to be made and in this work, we investigate the accuracy limits of ECPs of a simple semi-local form with almost a minimal number of parameters and we derive such ECPs for a small set of testing elements. 
As a guiding principle we have in mind {\it isospectrality for a subset of states}, i.e., we demand that the all-electron and ECP spectra are as close as possible to a set of valence states. 
The isospectrality 
is a very general property and applies even in cases when the Hilbert spaces for the two isospectral operators are different, e.g., due to different boundary conditions, different spatial domains, etc.

Note that when we refer to the spectrum, we have in mind the spectrum for not only the neutral atom, but states of cations, anions and binding curves for molecular systems containing this species.
In addition to spectra, we explore the use of many-body spatial information such as correlated single-body density matrices and natural orbitals as well as constructions that are matched to a combination of spatial and spectral information and further iterated constructions. 
These explorations help to identify trade-offs between the accuracy for spectra and spatial many-body properties versus demands on the constructions useful for a variety of electronic structure packages.
Additional demands such as systematic record of benchmarks from correlated calculations, appropriate  updates, etc., are further specified below.  

In Section \ref{desired}, we outline a list of properties that we argue are necessary for any ECP that leads to optimal results within high-accuracy valence-only many-body calculations. Section \ref{form} outlines the inverse-problem approach that we have utilized throughout this work as well as the form of our ECPs. Section \ref{optimization} covers general optimization methods and constructions for generating ECPs consistent with our desired set of properties. Results for the first- and second-row atoms, namely, boron, carbon, nitrogen, oxygen, and sulfur are shared in Section \ref{results}. An analysis of each pseudoatom's CCSD(T) dimer properties as compared to all-electron calculations is also given in Section \ref{results}. An extensive test of transferability is applied to our most accurate ECP constructions in Section \ref{transferability} in order to further ascertain their performance within various molecular environments. We conclude in Section \ref{conclusions}.

\section{Desired properties} \label{desired}
For this new generation of ECPs, we envision including the following set of properties.

a) {\bf Many-body construction} is the key and perhaps the most salient point in this list. Some ideas in this direction have been tried before,
for example, Acioli and Ceperley \cite{Acioli:1994jcp} explored correlated wave functions to generate ECPs and required that the pseudoatom's density matrix matched that of the full atom beyond a given core radius.
Extending on this methodology, Trail and Needs have generated tables of ECPs from MCHF wave functions \cite{Trail:2013jcp,Trail:2015jcp} and CCSD(T) excitation energies \cite{Trail:2017jcp} to improve their previously generated table \cite{Trail:2005jcp}. 
Additionally, attempts were made to build core effects into ECPs in the works of Maron, et al.\cite{Maron:cp1998} and Fromager et al.\cite{Fromager:jcp2004}.
We believe that a more systematic and accurate approach here can still be achieved, especially when considering properties outside of equilibrium (see below). 
In our construction, we use Coupled Cluster (CC) and Configuration Interaction (CI) methods that are very effective for atomic and small molecular systems, and moreover, we analyze how our generated ECPs perform in non-equilibrium geometries.

b) {\bf The simplicity of the ECP operator} has significant benefits since it allows use in many methods (QMC, quantum chemistry based on expansions in basis sets, and DFT, using established codes).
We, therefore, build upon a simple semi-local form that has been in use for some time. 
Per need, a number of extensions could prove to be useful and could be  incorporated in future. 
This might involve different choices of core-valence partitions, several types of representations such as numerical on radial meshes besides usual Gaussian expansions, fixed radial cutoffs, etc. 
More desired properties could be added as the project develops, e.g., how to optimally deal with ``fat" cores of heavy elements, possible inclusion of core polarization and core relaxation terms,  minimizing the fixed-node errors generated by the core part of the wave function without compromising 
the accuracy, minimizing the locality approximation in QMC applications following \cite{krogel_kent_2017}, etc. For plane wave applications, the suitability of recasting the ECP into the Kleinman-Bylander form \cite{Kleinman1982} could be included as well.

c) {\bf Testing} on a set of systems in order to delineate the accuracy limits for energy differences, equilibrium bond lengths and other properties. Potentially, some of these systems could be included into the retuning set, if important or necessary. For use and further improvements, it will be very useful to have documented systematic errors of generated ECPs documented upfront.

d) {\bf Systematic labeling and updates}, i.e., keeping the data, developments, and history on a website \cite{website} that can be eventually updated by interested contributors at large.

\section{Effective ECP Hamiltonian: Isospectrality on a subspace of valence states} \label{form}

One way to approach the ECP construction 
is to formulate it as  
an inverse problem, i.e.,
finding an effective operator that produces 
a subset of valence atomic states such that outside 
the core they reproduce the 
all-electron properties as closely as possible. 
We assume that for this subset both the all-electron and pseudized atoms can be solved in the same systematic and consistent
framework, ideally exactly, or, in practice, nearly exactly, using the state-of-the-art many-body methods.

What makes the two operators, the all-electron Hamiltonian and ECP Hamiltonian, close? Assuming we are interested in the valence subset of states and properties there are the following two key aspects:
     
\smallskip
\noindent
i) the two spectra should be the same/very close; and, 

\smallskip

\noindent
ii) the spatial characteristics (one- and multi-particle many-body density matrices) of the corresponding two sets of eigenstates outside the core region should be the same/very close. 

Note that, in general, the two Hamiltonians differ substantially, in the number of particles/degrees of freedom, presence/absence of ionic Coulomb singularities and relativity. 
Since in the reference calculations all correlations (including cores) are assumed to be present, i) and ii) implicitly demand that the ECPs capture, as much as possible, also the {\it impact of core-core and core-valence correlations
on the valence properties}. Our results below show that to a certain extent, this is indeed
the case.

Finding the desired solution in this setting essentially defines an inverse problem. We can expect that this problem will be ill-conditioned with many nearly optimal or non-unique solutions.
Why is this the case? The inclusion of many states into the optimization set often leads to a frustrated optimization problem, e.g., improvement for one state increases penalties for other states. With sufficient number of such frustrating couplings one ends up with a problem that is qualitatively similar, say, to finding a ground state of a spin glass (a well-known problem in statistical mechanics). 

One way how to deal with such 
ill-conditioned problems is to impose appropriate constraints 
that limit the space of possible solutions. That makes the problem solvable but could result in compromises on the 
resulting accuracy or incurred biases.
Often, it is quite difficult to find the right set of constraints. If the objective function is over-constrained the search for a minimum is fast but
the solution might be too biased by the constraints.
On the other hand, in an under-constrained formulation, the optimization could be very inefficient. 
Therefore, the goal is to find the best trade-off(s) between these two limits.

\subsection{ECP Form}

The valence-only electronic Born-Oppenheimer Hamiltonian we consider has 
the following form
\begin{equation}
   H_{\rm val} = \sum_i[T^{\rm kin}_i + V^{\rm pp}_i] +\sum_{i<j} 1/r_{ij}
\end{equation}
For this work, we use a semi-local ECP form with a minimal number of parameters \cite{Burkatzki:2007jcp}
\begin{equation}
    V^{\rm pp}_i = V_{loc}(r_i) + \sum_{l=0}^{l_{max}} V_l(r_i) \sum_{m}|lm\rangle\langle lm|,
\end{equation}
where $r_i$ is the radial distance of
electron $i$ from the core's origin. The non-local terms contain the projectors on $lm$-th angular momentum state.
The local term, $V_{loc}$, is chosen as
\begin{equation}
    V_{loc}(r) = -\frac{Z_{\rm eff}}{r}(1 -e^{-\alpha r^2}) + \alpha Z_{\rm eff} re^{-\beta r^2} + \gamma e^{-\delta r^2},
\end{equation}
where $Z_{\rm eff}$ is the effective core charge, $Z_{\rm eff}=Z-Z_{core}$.
The $V_l$ potential was chosen to consist of a single gaussian term
\begin{equation}
    V_l(r) = \beta_le^{-\alpha_lr^2}.
\end{equation}
All variables labeled by Greek letters are treated as optimization parameters in the minimization of a chosen objective function.
Additionally, a constraint that forces a concave shape at the origin is imposed \cite{Burkatzki:2007jcp}
\begin{equation}
    \gamma\delta + \alpha_l\beta_l > 0 , \quad \forall l .
\end{equation}
In the case(s) of combined constructions below, 
the number of gaussians in each channel is extended.

Note that the chosen form imposes a very significant restriction on the variational freedom of the effective operator. We have used this simple form not only to simplify
the optimization problem but also to investigate the true many-body accuracy limit for this ``minimal model" version.
The results are actually very encouraging. 
As presented in the following sections, even with such restricted variational freedom we were able to construct more accurate effective potentials than the existing ones as well as to illustrate the presented concepts above.   

\section{Optimization methods and constructions} \label{optimization}

In this section, we give a summary of a few atomic and molecular properties predicted from our constructed ECPs for several atoms from the first and second rows.
For comparison, along with our results, we juxtapose the predictions of these properties from correlated all-electron, uncorrelated core/correlated valence for a handful of other ECPs that have been used in many-body calculations in recent years.
Furthermore, we present results that utilize different strategies to construct the ECPs,
namely, ECPs built from all-electron spectral data only,
from spatial data only as well as combined and iterated constructions. 

For correlated calculations, a number of codes can be used in a non-relativistic setting. For relativistic calculations, however, the choices are more limited.
Nevertheless, it is interesting to see how far one could push 
the accuracy limits with currently existing 
methods and codes.

\subsection{Objective function with atomic spectral discrepancies only}

For the all-electron spectral-only references, we used the \textsc{Molpro} quantum chemistry package \cite{MOLPRO-WIREs} to calculate a subset of states from each atom's spectrum using the CCSD(T) method. 
To account for scalar relativistic effects, in the all-electron reference we used the Douglas-Kroll-Hess Hamiltonian throughout.
For each atom, the uncontracted aug-cc-pCV5Z basis set (which includes core state correlation functions) was chosen 
in order to minimize finite basis set errors as much as possible.
We used this same basis for both pseudo-atom and all-electron cases so that the basis set errors would largely cancel when gap discrepancies between the two cases were calculated. We carried out limited tests with even larger basis (6Z) and we found out that the gained accuracy had a marginal effect on the spectral differences when compared with other discrepancies
such as binding curves for molecular systems as explained below. In the supplementary material, we show again marginal level of errors from the extrapolation to complete basis set limit for both BFD and all-electron spectral gaps for the first and second row atoms N and S, respectively.

To generate the ECPs, we followed an energy consistent scheme, as described in reference \cite{Burkatzki:2007jcp} and references therein, and minimized the differences in the all-electron and ECP excitation energies in a least-squares way, with our objective function defined as 
\begin{equation}
    f = \sum_s \left( \Delta E_s^{AE} - \Delta E_s^{pp} \right)^2,
    \label{eqn:spectral}
\end{equation}
where $s$ labels a given excited state and $\Delta E_s$ is the energy gap between the excited state and the neutral atomic ground state.

We chose to include as many ionizations as possible in the atomic references; from the most deeply ionized single-valence state up to (at least) the $M^\mathrm{th}$ anionic state where $M$ is the predicted number of bound anions from the all-electron CCSD(T) spectrum.
The spectrum here therefore includes not only neutral atomic states, but also numerous states for the anions and cations.
In our investigation, we observed that when the most deeply ionized states were not included in the reference, their corresponding excitations tended to have significant discrepancies from the all-electron excitations and consequently the pseudoatom's transferability could be negatively affected.
Additionally, for each ionization, we expanded the reference to include the bound ground states of all possible total spin channels of that ion, e.g., for an ion with four electrons in the valence space, we included the ground states of the quintet, triplet and singlet spin channels into the reference, provided that each state was predicted to be bound from the all-electron calculation.
The latter choice was motivated by a desire to minimize any possible contamination from energetically lower states within the same symmetry channel.

\subsection{ Objective function with spectral and spatial density matrix discrepancies}
In order to incorporate spatial information into the pseudopotential, we explored the utilization of the single-body density matrix following the previous work of Acioli and Ceperley. 
For the all-electron reference and pseudoatom, we generated the single-body density matrices from a CISD wave function.
We expect that if the pseudoatom density matrix and the all-electron density matrix agree beyond a chosen core region defined by
cut-off radius $r_c$, then the pseudoatom valence will mimic that of the all-electron atom.
Given an AE atom and a pseudoatom, we can measure the
difference between the density matrices outside of $r_c$ as
\begin{equation}
    \Delta\rho_{r_c} = \iint\limits_{r,r'>r_c} d\mathbf{r} d\mathbf{r}' |\rho_{AE}(\mathbf{r},\mathbf{r}') - \rho_{ECP}(\mathbf{r},\mathbf{r}')|^2 .
    \label{eqn:dm}
\end{equation}
If we express each density matrix through the natural orbitals $\{\phi_i\}$ and occupation numbers $n_i$ of a correlated calculation such as CISD, we can rewrite the discrepancy between the density matrices as
\begin{eqnarray}
    \Delta\rho_{r_c} = \sum_{ij}^N n^{AE}_i n^{AE}_j \left|\langle \phi_i^{AE} | \phi_j^{AE}\rangle_{r_c}\right|^2 
    \nonumber\\
    + \sum_{i,j}^M n_i^{ECP}n_j^{ECP} \left| \langle \phi_i^{ECP}|\phi_j^{ECP}\rangle_{r_c}\right|^2 \nonumber\\
    -2\sum_i^N \sum_j^M n_i^{AE} n_j^{ECP} \left| \langle\phi_i^{AE} | \phi_j^{ECP} \rangle_{r_c}\right|^2
\end{eqnarray}
where the expression $\langle \phi_i | \phi_j \rangle_{r_c}$ is the overlap of two natural orbitals evaluated from $r_c$ to infinity.

An important point worth noting here is that  
one should not expect to rigorously match the all-electron and pseudoatom density matrices. 
The reason is that the density matrix of the original 
atom generally contains a contribution from the tails 
of the core states beyond the cutoff radius. Since it is difficult to disentangle the tails in the many-body setting this ``contamination" will be present.
Note that in traditional one-particle formulations this is a non-issue since 
HF or Kohn-Sham self-consistent orbitals  
decompose 
the core and valence states into distinct sets. 
The closest analogy to the 
one-particle decomposition comes from the natural orbitals, however, even there the occupation numbers 
and the natural orbitals will slightly differ 
 (reflecting thus the presence or the absence of the core states).
 This becomes particularly relevant whenever smaller core radii are imposed. 
The fits can be further elaborated to approximately take this into 
account, however, here we opted for a simple agreement
assuming that this effect will not dominate the achieved
discrepancies.

We further constructed ECPs where we include both the spectral and spatial information into the objective function. 
Using the integration of the density matrix beyond a cutoff radius as defined above, we define our spatial objective function as
\begin{equation}
    g = \sum_s \Delta \rho_{r_c}^s
    \label{eqn:spatial}
\end{equation}
where we sum over all states considered.
We construct a new objective function where we include both spatial and spectral information as
\begin{equation}
    h = \frac{f}{f_0} + \frac{g}{g_0}
    \label{eqn:spec/space}
\end{equation}
where $f_0$ and $g_0$ are the optimal values of the respective objective functions. In the case of sulfur below, the two objective functions were left unweighted given the similarity between $f_0$ and $g_0$ in that case.

\subsection{Optimization methods}
Throughout, we have utilized the nonlinear DONLP2 optimization code of Spellucci \cite{DONLP2} for generating the final parameter values of our ECPs with respect to the various objective functions which we have considered.
The specifics of the method implemented in DONLP2 are outlined in two papers \cite{SpellucciA,SpellucciB}, where generally, the solver extends the sequential quadratic programming method, an iterative method that relies on the second derivatives of both the objective function and the constraints, such that it can be applied to nonlinear problems.
We considered this choice for the chosen solver to be appropriate given the nonlinear ``smoothness" constraint that has been imposed on the parameter sets of our ECPs.

\subsection{ Constructed, Combined and Iterated Schemes}

The final procedure that we pursued took molecular data into account.
Having generated molecular binding curve predictions from an assortment of ECP constructions, we noticed that there was a potential to improve the ECPs further. In particular, the optimization could be guided to fulfill additional criteria such as reproducing the dimer binding curve in a few iterations. This type of optimization loop produced very small discrepancies from the all-electron CCSD(T) curves along the desired range of geometries even at the steep repulsive side of the binding curves. 
In particular, for the corresponding dimers, we found that this strategy is able to produce discrepancies as small as $0.05$~eV or lower along the binding curve all the way up to the dissociation limit at the repulsive side. Due to time demands of correlated calculations and rather slow coupling between different codes we guided this level of optimization by interventions to speed up the search. That was also very useful to understand the qualitative relationships between shapes/amplitudes of the ECP versus its properties and therefore we leave the automation of this part for future. We comment on these constructions at each presented atomic case.

In all of the systems presented below, we compare our ECPs to all-electron CCSD(T), all-electron CCSD(T) with an uncorrelated core (UC), i.e., no excitations from the core, and various ECPs often used in QMC calculations, namely Burkatzki-Filippi-Dolg (BFD) \cite{Burkatzki:2007jcp}, Trail-Needs (TN-DF) \cite{Trail:2005jcp} DF ECPs, and where applicable, the shape-consistent correlated  Trail-Needs  (TN-CEPP) ECPs \cite{Trail:2013jcp} as well as the recent shape and energy consistent correlated Trail-Needs (TN-eCEPP) ECPs \cite{Trail:2017jcp}. Note that when utilizing the TN-CEPP and TN-eCEPP, we only test the effective core potential and do not include the semi-empirical core polarization potentials utilized by Trail and Needs, 
given that we were interested in comparing all effective cores within a consistent level of approximation. 
A key question that we chose to pursue was how well an ECP alone would capture core-valence effects without the need for additional adjustments or approximations. By not including CPPs along with the ECPs allowed us to isolate this effect.
Note that, Trail and Needs have studied the effect of the CPP in Figure 6 of \cite{Trail:2017jcp}, illustrating that CPPs do have only minor impact on the dissociation energies over a wide variety of molecules. As explained in results, we found that much larger discrepancies almost invariably appear at shorter molecular bond lengths and therefore we have focused 
on addressing this aspect in
our constructions. 

\section{Results} \label{results}
Here we present our initial results for a number of explicitly correlated ECPs using the various objective functions described above. In the first row, we present optimized ECPs for Boron, Carbon, Nitrogen, and Oxygen. For the second row elements, we show our results for Sulfur.
\subsection{Boron}
\begin{table*}[ht!]
        \centering
        \caption{ Atomic and ionic excitations and corresponding discrepancies for Boron. IP denotes the first ionization potential while EA is the electron affinity. Q is the ionization charge, 2S+1 the usual total spin multiplicity. AE denotes the calculated all-electron values while the rest of columns shows the discrepancies. 
        UC means all-electron valence-only correlation with self-consistent but uncorrelated core, as explained in the text.
        All energies in eV. The MAD is the mean absolute difference over all of the discrepancies. 
        Note that all gaps are calculated with reference to the ground state, namely Q=0 and 2S+1 = 2.
        The same notation applies to all the atomic/ionic data tables throughout the paper. }
        \begin{tabular}{rcrrrrrrrr}
              \hline
              \hline
              Q &   2S+1                  & AE Gap      & UC      & BFD     & TN-DF    & TN-CEPP & TN-eCEPP &  Spectral & Constructed \\
               \hline
             +2 & 2 & $ 33.4290$  & $ -0.0726 $ & $  0.0048 $ & $  -0.0316 $ & $   0.1186 $  & $  0.0224 $   &  $   0.0027 $ &  $  0.0109 $ \\  
         (IP)+1 & 1 & $  8.2771$  & $ -0.0379 $ & $ -0.0670 $ & $  -0.0711 $ & $  -0.0030 $   & $ -0.0280 $   &  $  -0.0077 $ &  $ -0.0301 $ \\
             +1 & 3 & $ 12.9083$  & $ -0.0085 $ & $  0.0249 $ & $   0.0121 $ & $   0.0326 $  & $  0.0032 $   &  $  -0.0042 $ &  $ -0.0129 $ \\
              0 & 4 & $  3.5752$  & $  0.0187 $ & $  0.0202 $ & $   0.0523 $ & $   0.0184 $  & $  0.0107 $   &  $   0.0018 $ &  $ -0.0079 $ \\
           $-$1 & 1 & $  0.2854$  & $ -0.0026 $ & $  0.0067 $ & $   0.0092 $ & $   0.0079 $  & $  0.0046 $   &  $  -0.0034 $ &  $  0.0031 $ \\
      (-EA)$-$1 & 3 & $ -0.2482$  & $  0.0043 $ & $  0.0142 $ & $   0.0220 $ & $   0.0144 $ & $  0.0117 $   &  $   0.0006 $ &  $  0.0098 $ \\
           $-$1 & 5 & $  2.5144$  & $  0.0207 $ & $  0.0183 $ & $   0.0683 $ & $   0.0301 $  & $  0.0193 $   &  $  -0.0004 $ &  $ -0.0027 $ \\
              \hline
              MAD & & & 0.0236 & 0.0223 & 0.0381 & 0.0321 & 0.0143 & 0.0030 & 0.0110\\
              \hline
              \hline
           \end{tabular}
        \label{tab:boron}
    \end{table*}

\begin{figure}[ht!]
    \caption{Boron dimer potential energy surface. UC represents an all-electron CCSD(T) calculation with a self-consistent but {\it uncorrelated} core, i.e., with
no excitations from the core states. Spectral represents the optimization for the atomic spectrum alone, and Constructed indicates the ECP driven iteratively to minimize the dimer discrepancy while accepting a small increase in the spectrum discrepancy.}
\includegraphics[width=8cm]{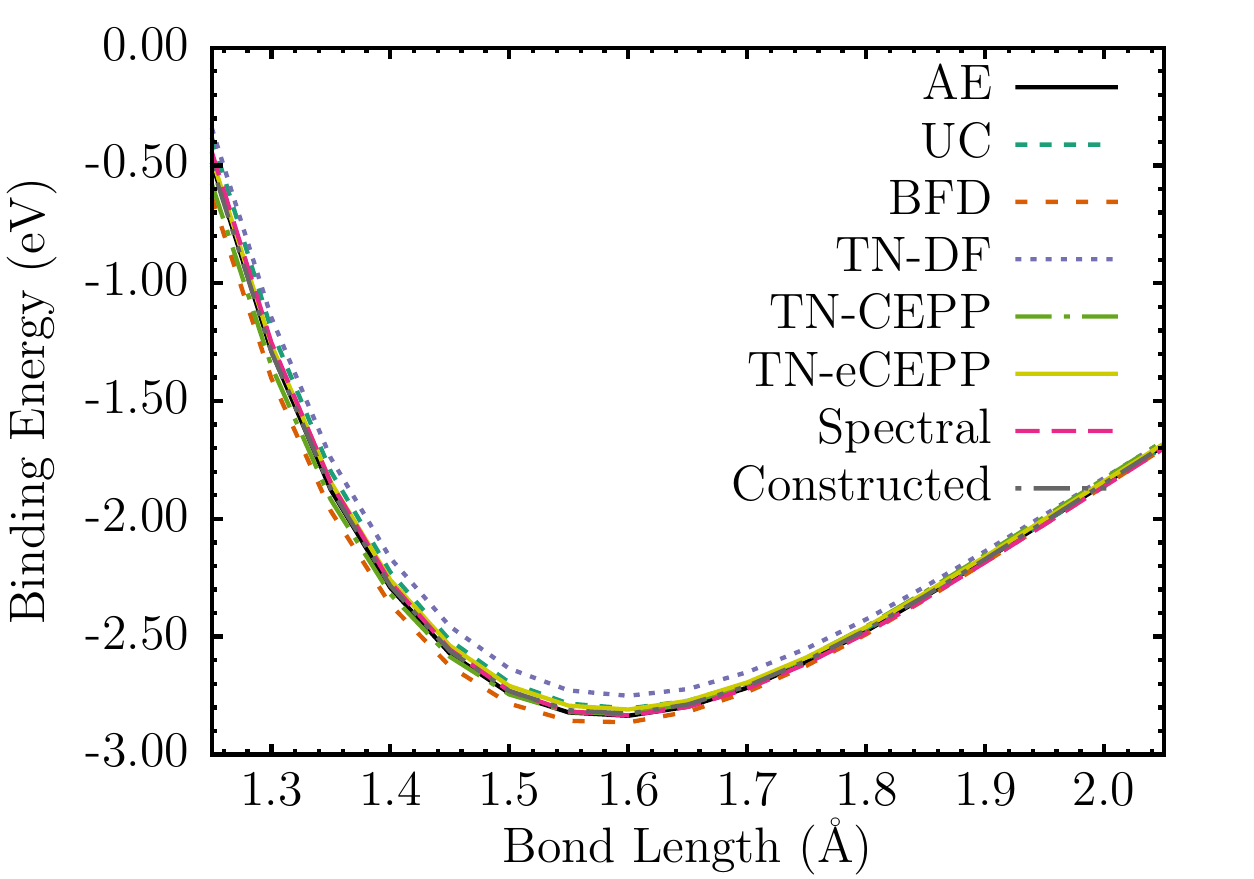}
\label{fig:b2}
\centering
\end{figure}

\begin{figure}[ht!]
    \caption{Boron dimer binding energy discrepancies compared to the all-electron CCSD(T) binding curve. The gray envelope represents a $0.05$~eV window for the discrepancy. The vertical line indicates the equilibrium bond length as predicted by the all-electron CCSD(T) calculation. }
\includegraphics[width=8cm]{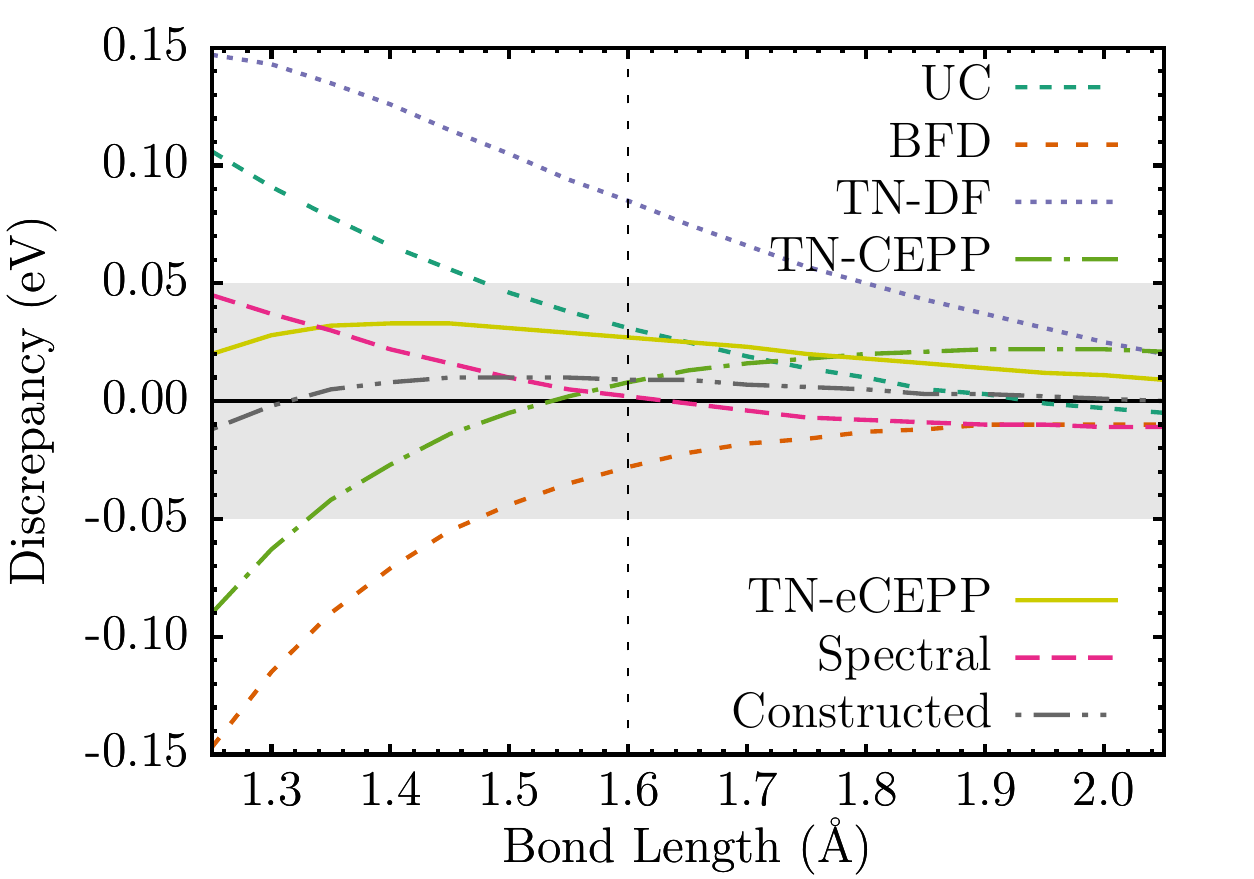}
\label{fig:b2_diffs}
\centering
\end{figure}

Boron is perhaps the lightest element where time savings from an appropriate ECP become significant while the valence space is sufficiently large to represent the atom in chemical settings. 
Table \ref{tab:boron} shows 
several constructions of the ECP  obtained by minimizing the atomic spectral error only (Spectral) and another that has been adjusted to reproduce the binding 
curve of the ground state of B$_2$ in the range of bond lengths (Constructed).
Note the very high accuracy that was obtained with minimizing only the spectral discrepancy. A significantly improved dimer solution has been 
found by minimizing a compromise between reproducing the spectrum
and the binding curve. Note that the shortest 
bond length corresponds approximately to the dissociation point due to nucleus-nucleus repulsion. In solids, such distance between
the atoms would correspond to very high pressures roughly beyond 
500~GPa. 
Figure \ref{fig:b2} shows the B$_2$ potential energy surface for the $^3\Sigma_g$ state and \ref{fig:b2_diffs} provides a set of discrepancies from the all-electron CCSD(T) for our constructed ECPs  compared against
previous ECPs: BFD, TN-DF, the recently correlated constructions by Trail and Needs (TN-CEPP and TN-eCEPP) as well as the all-electron uncorrelated core result.

\begin{table}[!ht]
    \setlength{\tabcolsep}{12pt}
    \centering
    \caption{ECP parameters for Constructed Boron. The parametrization for each channel is given by $V_l(r) = \sum_k \beta_{lk} r^{n_{lk}-2}e^{-\alpha_{lk}r^2}$. The corresponding correlation consistent basis sets are included in the Supplementary Material. }
    \begin{tabular}{c r r r}
	\hline\hline
	Channel & $n_{lk}$ & $\alpha_{lk}$  & $\beta_{lk}$ \\
	\hline
	$p$     &  1  & 31.49298 &  3.00000\\
	        &  3  & 22.56509 & 94.47895\\
		    &  2  &  8.64669 & -9.74800\\
	\hline
	$s-p$   &  2  & 4.06246  & 20.74800\\
	\hline\hline
    \end{tabular}
    \label{tab:pp_b}
\end{table}

Right away we can observe the following:

a) The constructed ECPs provide a significantly more accurate picture of the molecular system, both for the valence spectrum optimization (Spectral) as well
as more accurate binding curve optimization (Constructed), than the existing examples. Remarkably, our ECPs show 
{\it smaller errors} for both the spectrum and for binding curve than the all-electron correlated valence-valence calculation with an {\it uncorrelated 
core} (UC). In UC all one-particle states are solved self-consistently including the cores (i.e., it is not the atomic frozen core in that sense), however, only valence-valence correlations are invoked in the CCSD(T) method while core-valence and core-core correlations are neglected. The fact that our effective Hamiltonians provide a better description than UC for the dimer and atomic spectra is quite surprising. We, therefore, observe that the inverse problem formulation enables to mimic, within a certain level of accuracy, effects of core excitations on the valence space. 

b) The simple form of the ECP described above is able to accommodate higher accuracy demands.  

Note that overall accuracy in reproducing correlated 
properties is roughly at the level of $0.02$~eV, significantly higher than 
in previous constructions.
Both of these observations are quite unexpected and will be further
elaborated in the cases of other elements.
Since these solutions were satisfactory we have not pursued further
improvements by including spatial correlations explicitly into
the objective function. Note that the spectrum and refinement
alone were sufficient to determine the high accuracy solution
within the minimal and significantly restricted representation.
Our best ECP parameters (Constructed) are given in Table \ref{tab:pp_b}. 

{\bf Labeling.}
For the sake of clear identification and recording of progress we introduce here the label for this and subsequent ECPs as ``correlation consistent ECP",
version 0.1, or {\bf ccECP.0.1}, in short.
The same label is assigned to all tabulated ECPs in this paper.

Note that the notation can be further expanded, for example, as ttECP.xx.yy.ENSCO.
 Here the theory would be labeled as
tt=cc,hf,df,..., while major.minor release is represented as above by xx.yy, respectively. In addition, fitted quantities could include ENSCO=excitations-norms/shapes-spatialDM-combined/iterated-other approaches. As far as systems used for fit is concerned one could possibly affix a label ADHOCG=atom-dimer-hydride-oxide-cluster-general/other system. As an example, the full label for our Boron case 
would be ccECP.0.1.EC.

\subsection{Carbon}

\begin{figure}[ht!]
\caption{Carbon dimer binding energy discrepancies compared to the all-electron CCSD(T) binding curve.}
\includegraphics[width=8cm]{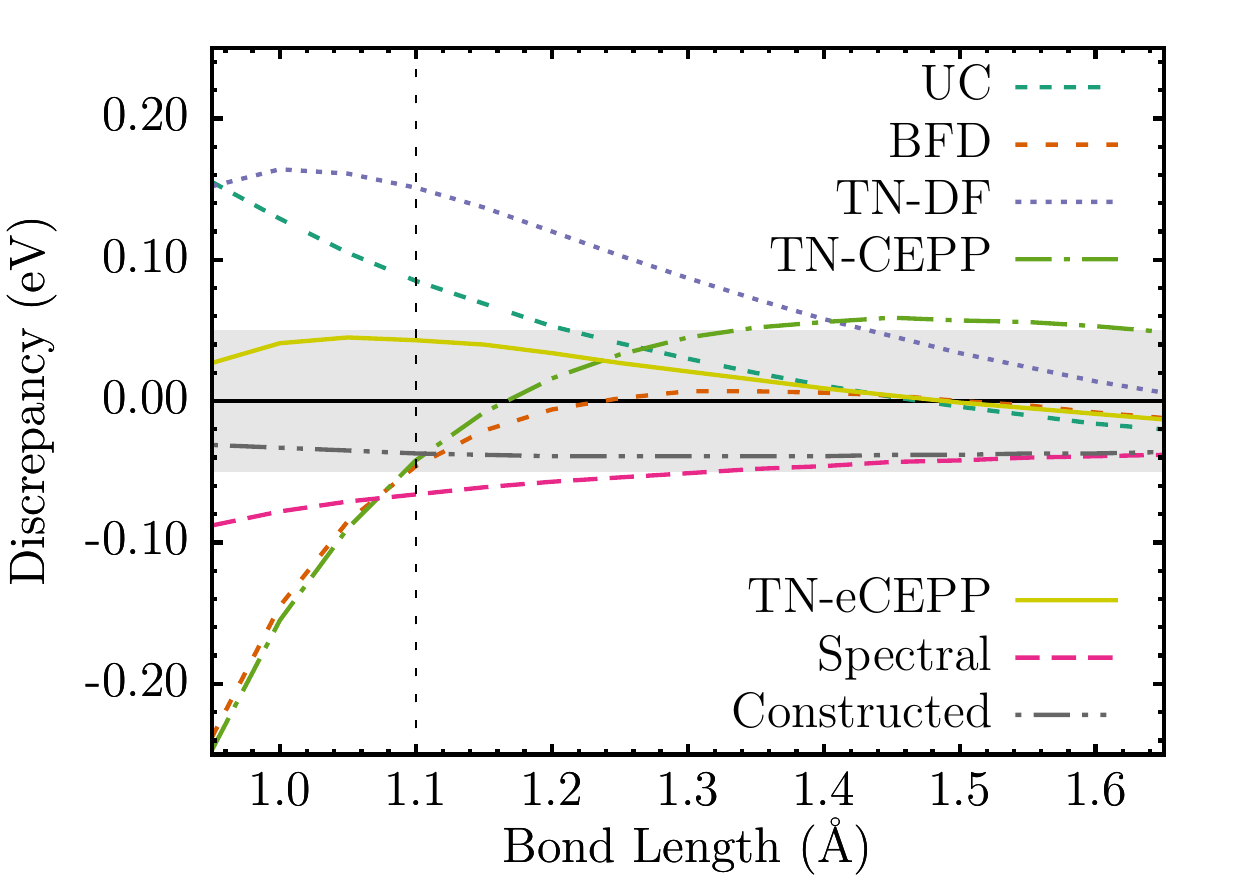}
\label{fig:c2_all_diffs}
\centering
\end{figure}

Since the spectral only optimization worked quite well for B, only needing a slight readjustment with our constructed ECP, we test the same for the C atom.
The atomic spectrum for various ECPs is given in Table \ref{tab:carbon} and dimer discrepancies in the $^1\Sigma_g$ state are given in Figure \ref{fig:c2_all_diffs}.
The spectral only construction does remarkably well on minimizing the energy differences for the atomic spectra and produces an ECP that is improved over other ECPs. While there is slight overbinding, the Spectral curve is relatively flat indicating accurate vibrational frequencies compared to the all-electron.  The electron affinity (EA) for the spectral differs with the all electron CCSD(T) value by only $-1$~meV. The ionization potential (IP) is an order of magnitude better and only differs by $+0.1$~meV. 
In addition, a minor compromise in reproducing the atomic spectra produces a much better dimer binding that is within $0.05$~eV across the entire curve and has a flat discrepancy, i.e., the same vibrational frequencies as the all-electron curve. 
\begin{table*}[ht!]
    \centering
    \caption{Atomic data for Carbon, similar to Table \ref{tab:boron}. Energies in eV. Note that all gaps are calculated with reference to the ground state, namely Q=0 and 2S+1 = 3.}
    \begin{tabular}{rcrrrrrrrr}
	\hline
	\hline
	Q &  2S+1   & AE Gap   &       UC      &       BFD     &     TN-DF    &   TN-CEPP    &   TN-eCEPP   &     Spectral & Constructed \\
	\hline
	$+3$ & 2 & $ 83.4895 $ & $ -0.1469 $ &  $ -0.1090 $ & $ -0.1561 $ & $ -0.0824 $  & $  0.2544 $ & $  0.0005 $ & $ -0.0024 $\\  
	$+2$ & 1 & $ 35.6041 $ & $ -0.1020 $ &  $ -0.2208 $ & $ -0.1723 $ & $ -0.0953 $  & $  0.0326 $ & $ -0.0007 $ & $  0.0110 $\\
	$+2$ & 3 & $ 42.1035 $ & $ -0.0561 $ &  $ -0.1083 $ & $ -0.1080 $ & $ -0.0368 $  & $  0.0751 $ & $ -0.0009 $ & $ -0.0061 $\\
    (IP)$+1$ & 2 & $ 11.2452 $ & $ -0.0334 $ &  $ -0.0725 $ & $ -0.0631 $ & $ -0.0277 $  & $ -0.0073 $ & $  0.0016 $ & $  0.0027 $\\
        $+1$ & 4 & $ 16.5590 $ & $ -0.0022 $ &  $ -0.0955 $ & $ -0.0552 $ & $ -0.0173 $  & $  0.0075 $ & $  0.0001 $ & $  0.0019 $\\
	 $0$ & 1 & $  1.3950 $ & $ -0.0143 $ &  $  0.0013 $ & $  0.0000 $ & $  0.0045 $  & $  0.0095 $ & $  0.0006 $ & $ -0.0009 $\\
	 $0$ & 5 & $  4.1491 $ & $  0.0231 $ &  $ -0.0743 $ & $ -0.0085 $ & $ -0.0030 $  & $  0.0126 $ & $  0.0005 $ & $  0.0084 $\\
   (-EA)$-1$ & 4 & $ -1.2421 $ & $  0.0072 $ &  $  0.0259 $ & $  0.0249 $ & $  0.0098 $  & $  0.0270 $ & $ -0.0010 $ & $ -0.0006 $\\
	 \hline
	 MAD &   &              &    0.0481      &    0.0884    &    0.0735    &    0.0532    &   0.0346     &     0.0008  &    0.0046 \\
	 \hline
	 \hline
    \end{tabular}
    \label{tab:carbon}
\end{table*}

The parameters for our best Carbon ECP (Constructed) is given in Table \ref{tab:pp_c}.

\begin{table}[h]
    \setlength{\tabcolsep}{12pt}
    \centering
    \caption{ECP parameters for Constructed Carbon. The parametrization for each channel is given by $V_l(r) = \sum_k \beta_{lk} r^{n_{lk}-2}e^{-\alpha_{lk}r^2}$. The corresponding correlation consistent basis sets are included in the Supplementary Material. }
    \begin{tabular}{c r r r}
	\hline\hline
	Channel & $n_{lk}$ & $\alpha_{lk}$  & $\beta_{lk}$ \\
	\hline
	$p$     &  1 & 14.43502 & 4.00000 \\
	        &  3 & 8.39889  & 57.74008 \\
		&  2 & 7.38188  & -25.81955 \\
		\hline
	$s-p$   &  2 & 7.76079  & 52.13345  \\
	\hline\hline
    \end{tabular}
    \label{tab:pp_c}
\end{table}

\subsection{Nitrogen}

\begin{table*}[ht!]
    \centering
    \caption{Atomic data for Nitrogen, similar to Table \ref{tab:boron}. Energies in eV. Note that all gaps are calculated with reference to the ground state, namely Q=0 and 2S+1 = 4.}
    \scriptsize
    \begin{tabular}{rcrrrrrrrrrr}
	\hline
	\hline
	Q &   2S+1         & AE Gap      & UC    &    BFD     &   TN         &  TN-CEPP   & TN-eCEPP    &    Spectral   & Spatial    & Spec/Space & Constructed \\
	\hline
     	     $+4$ & $2$ & $169.0094$ & $-0.2308$ & $ -0.2139$ & $ -0.3301$   & $ 0.5136 $ &  $-0.3450 $ &   $  0.0024 $ & $ -0.2259 $& $  0.0022 $ &  $ -0.2758 $ \\   
     	     $+3$ & $1$ & $ 91.5371$ & $-0.1784$ & $ -0.3903$ & $ -0.2882$   & $ 0.2012 $ &  $-0.1921 $ &   $ -0.0043 $ & $ -0.2368 $& $ -0.0046 $ &  $ -0.0494 $ \\
     	     $+3$ & $3$ & $ 99.8869$ & $-0.1205$ & $ -0.2582$ & $ -0.2868$   & $ 0.2291 $ &  $-0.2538 $ &   $ -0.0011 $ & $ -0.0157 $& $ -0.0009 $ &  $ -0.2531 $ \\
     	     $+2$ & $2$ & $ 44.1212$ & $-0.0866$ & $ -0.1827$ & $ -0.1429$   & $ 0.0792 $ &  $-0.0862 $ &   $  0.0018 $ & $ -0.0635 $& $  0.0017 $ &  $  0.0014 $ \\
     	     $+2$ & $4$ & $ 51.1908$ & $-0.0455$ & $ -0.2695$ & $ -0.2339$   & $ 0.0645 $ &  $-0.1732 $ &   $ -0.0023 $ & $  0.0548 $& $ -0.0019 $ &  $ -0.2317 $ \\
     	     $+1$ & $1$ & $ 16.5790$ & $-0.0448$ & $ -0.0472$ & $ -0.0370$   & $ 0.0380 $ &  $-0.0168 $ &   $  0.0105 $ & $  0.0028 $& $  0.0105 $ &  $  0.0232 $ \\
         (IP)$+1$ & $3$ & $ 14.5319$ & $-0.0297$ & $ -0.0646$ & $ -0.0547$   & $ 0.0125 $ &  $-0.0291 $ &   $ -0.0013 $ & $ -0.0070 $& $ -0.0013 $ &  $  0.0100 $ \\
  	     $+1$ & $5$ & $ 20.5319$ & $ 0.0009$ & $ -0.2524$ & $ -0.1729$   & $-0.0041 $ &  $-0.1077 $ &   $  0.0011 $ & $  0.0559 $& $  0.0015 $ &  $ -0.2154 $ \\
  	     $ 0$ & $2$ & $ 2.6789 $ & $-0.0295$ & $ -0.0037$ & $ -0.0056$   & $ 0.0046 $ &  $-0.0104 $ &   $ -0.0084 $ & $ -0.0080 $& $ -0.0085 $ &  $ -0.0068 $ \\
	 \hline
	 MAD &   &                   &   0.0852  &   $0.1870$  &  0.1725      & 0.1274     &   0.1349    &      0.0037      & 0.0745 & 0.0037  &   0.1185\\
	 \hline
	 \hline
    \end{tabular}
    \label{tab:nitrogen}
\end{table*}

\begin{figure}[ht!]
\caption{Nitrogen dimer binding energy discrepancies compared to the all-electron CCSD(T) binding curve.}
\includegraphics[width=8cm]{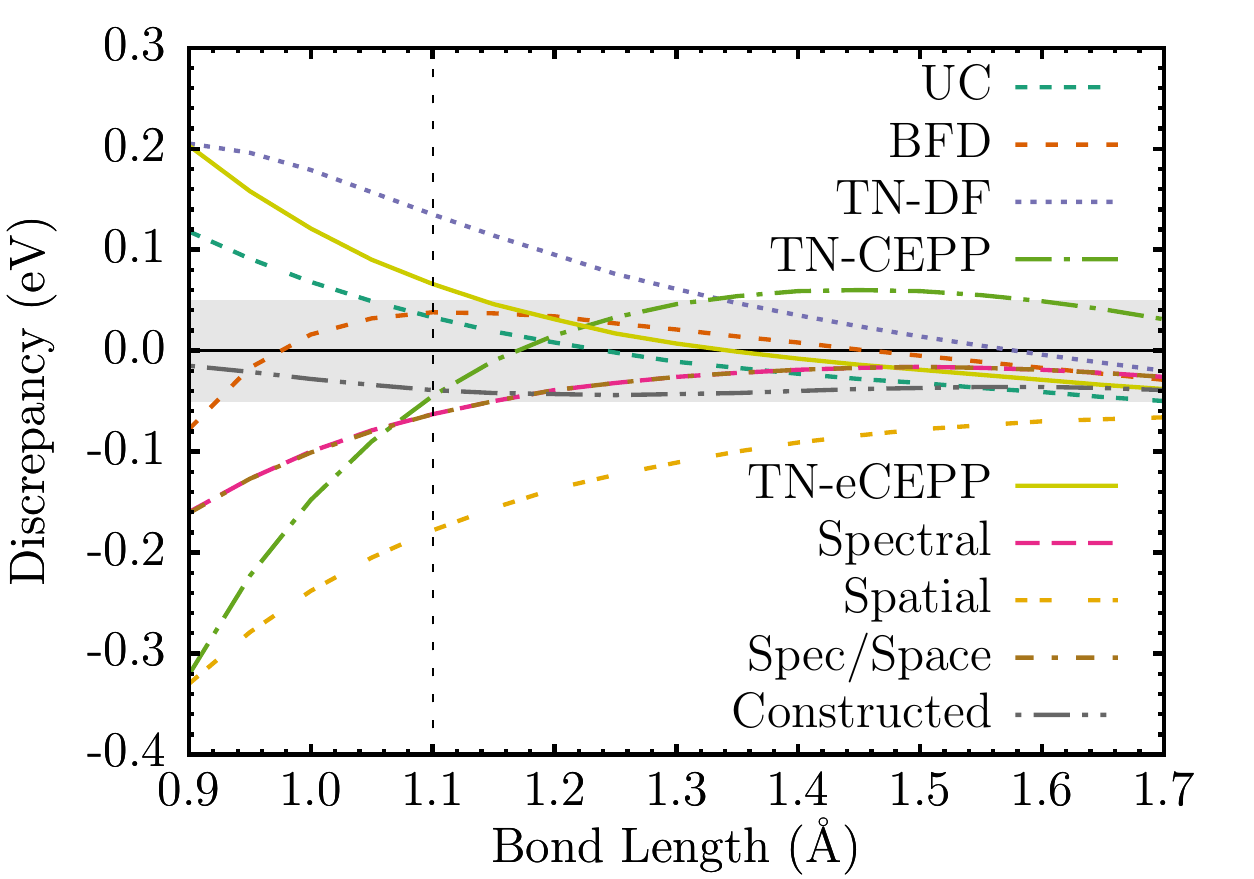}
\label{fig:n2_all_diffs}
\centering
\end{figure}

In constructions of nitrogen, we explored several objective functions and refinement strategies, as shown in Figure \ref{fig:n2_all_diffs} for the molecule in the $^1\Sigma_g$ state and Table \ref{tab:nitrogen} for the atomic properties.
The spectrum only optimization performs rather well and is a significant improvement over many of the previous constructions. 
The results show that using the spatial information only
such as one-particle density matrices appears 
less favorable overall. We conjecture that this
is caused due to very weak correlation ``signal" in the corresponding density matrices. The difference 
between the mean-field and correlated density matrices is very tiny so that the correlation effect is overwhelmed by the one-particle character that is dominant.
On the other hand, the spectrum alone 
provides a much stronger signal since the eigenvalues basically determine the wave function tails essentially exactly. 
When combining the spatial and spectral information, the spatial signal was not sufficient to significantly change the parameters, and we obtain essentially the same ECP for both constructions. 
Although the spectral ECP has excellent atomic properties and a reasonable dimer, we sought to construct a better ECP overall. 
The constructed ECP was generated by adding additional constraints that were not included in the original objective function in order to alter systematic trends we observed in our various constructions to minimize the dimer discrepancy. 
In particular, we observed that pairs of ECPs with dimer discrepancies of opposite trends could be produced based on different constraints and subsequently the ECPs could be combined as a linear sum leading to increased agreement with the molecular properties from all-electron CCSD(T) while also preserving a high level of accuracy on the atomic spectrum.
This procedure could certainly be used to define a new objective function where the dimer fit is directly included, however, this proved to be computationally inefficient and we therefore chose to improve the molecular properties by hand as described.
The constructed ECP has a discrepancy of below $0.05$~eV across the entire curve, although we compromise the spectrum when compared to our spectral construction, particularly for high-spin atomic states. 
However, if the high-spin atomic states are not considered, the atomic spectrum is better than all previous generations of ECPs. Note that in molecular systems the nitrogen almost invariably appears in low/lowest spin configurations.

Another observation is that ECPs with a combination of minimizing the dimer binding and spectrum discrepancies could be constructed without generating an overly large negative impact to the atomic spectra, provided an appropriate constraint is utilized.  
Though some accuracy of the atomic properties would diminish as the dimer binding is improved, as can be seen in Table \ref{tab:nitrogen}, we found that a reasonable balance could be obtained between the two when compared to other ECP constructions. 

The parameters for our best Nitrogen ECP (Constructed) is given in Table \ref{tab:pp_n}.

\begin{table}[h]
    \setlength{\tabcolsep}{12pt}
    \centering
    \caption{ECP parameters for Constructed Nitrogen. The parametrization for each channel is given by $V_l(r) = \sum_k \beta_{lk} r^{n_{lk}-2}e^{-\alpha_{lk}r^2}$.
    The corresponding correlation consistent basis sets are included in the Supplementary Material. }
    \begin{tabular}{c r r r}
	\hline\hline
	Channel & $n_{lk}$ & $\alpha_{lk}$  & $\beta_{lk}$ \\
	\hline
	$p$     &   1      & 12.91881    &   3.25000 \\
	        &   1      &  9.22825    &   1.75000 \\
		&   3      & 12.96581    &  41.98612 \\
		&   3      &  8.05477    &  16.14945 \\
		&   2      & 12.54876    & -26.09522 \\
		&   2      &  7.53360    & -10.32626 \\
		\hline
	$s-p$   &   2      &  9.41609    &  34.77692 \\
	        &   2      &  8.16694    &  15.20330 \\
	\hline\hline
    \end{tabular}
    \label{tab:pp_n}
\end{table}

\subsection{Oxygen}
\begin{figure}[ht!]
\caption{Oxygen dimer binding energy discrepancies compared to the all-electron CCSD(T) binding curve.}
\includegraphics[width=8cm]{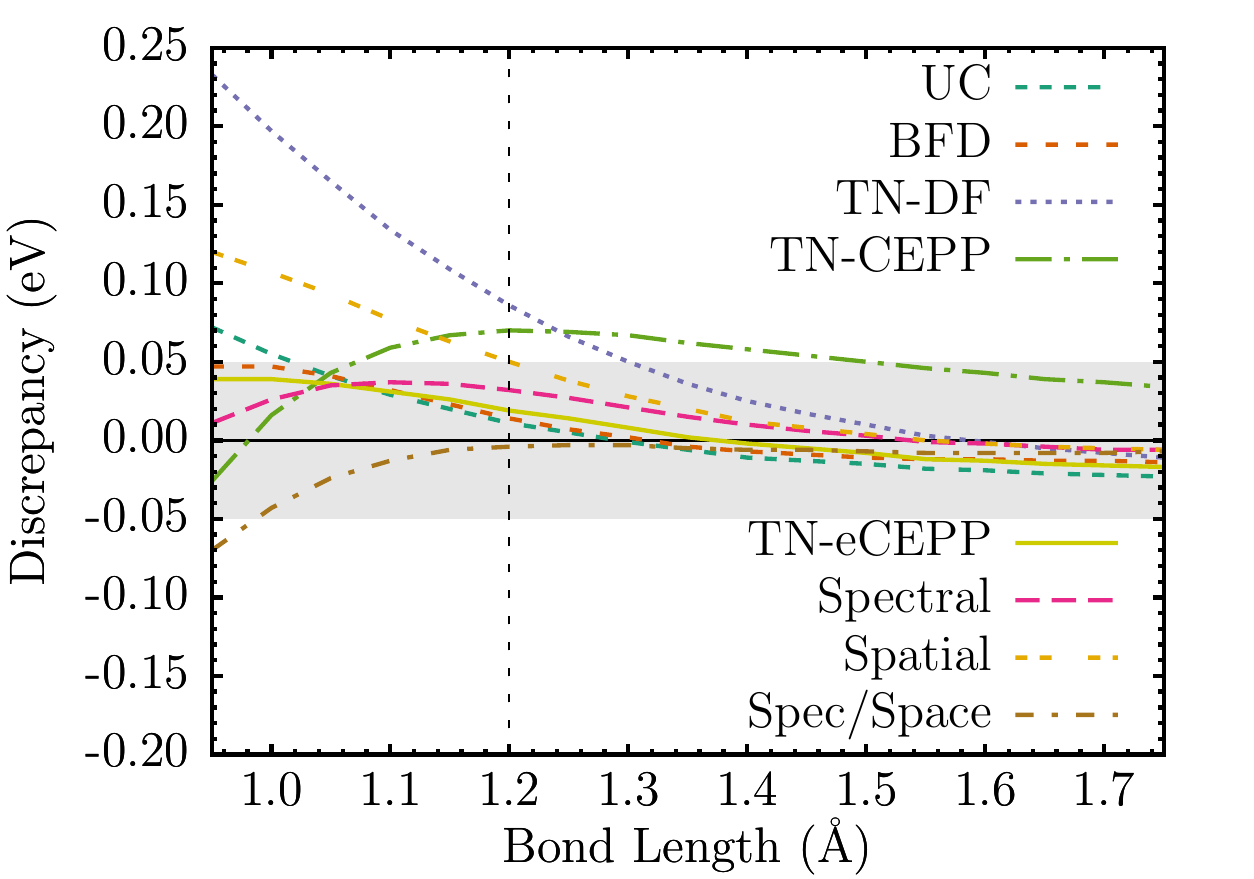}
\label{fig:o2_all_diffs}
\centering
\end{figure}

As in the previous case, we test a variety of objective functions when constructing an ECP for O. 
We present the results in Figure \ref{fig:o2_all_diffs} for the molecule in the $^3\Sigma_g$ state  and in Table \ref{tab:oxygen} for the atomic properties.
Using a spectral-only construction, we are able to construct an ECP that provides a binding curve discrepancy to the all-electron CCSD(T) curve within $0.05$~eV along the entire curve. 
Since the Spectral ECPs are reasonably flat compared to other ECP constructions and the all-electron uncorrelated core (UC), the vibrational frequencies agree quite well to the all-electron CCSD(T) calculations. As in N, the spatial information alone appears insufficient in generating a high-quality ECP. 
When combining both spatial and spectral information (Spec/Space), we actually obtain an ECP that is essentially flat across a wide range of bond lengths and shows the best dissociation energy. This also results in a slight compromise on the atomic properties. 
However, this ECP begins to overbind significantly in the shorter bond length regime which would correspond to high pressure in solids. Based on this and the quality of our spectral only ECP, we believe that the spectral only optimization for Oxygen produces the best ECP overall and we do not pursue iterated constructions. The parameters for Spectral Oxygen are given in Table \ref{tab:pp_o}

\begin{table*}[ht!]
    \centering
    \caption{Atomic data for Oxygen, similar to Table I. Energies in eV.  Note that all gaps are calculated with reference to the ground state, namely Q=0 and 2S+1 = 3.}
    \scriptsize
    \begin{tabular}{rcrrrrrrrrr}
	\hline
	\hline
	Q &  2S+1   & AE Gap   &       UC      &    BFD  &      TN-DF    & TN-CEPP  & TN-eCEPP  & Spectral & Spatial & Spec/Space \\
	\hline
      $+5$ & $2$  & $294.8903$ & $ -0.3059$ & $ -0.4334$ & $ -0.6186$ & $ 0.7414$   &  $-0.5643 $ & $-0.0021$ & $ -0.4531$ & $ -0.0228$ \\  
      $+4$ & $1$  & $180.9900$ & $ -0.2478$ & $ -0.6864$ & $ -0.4477$ & $ 0.3505$   &  $-0.1826 $ & $ 0.0017$ & $ -0.2573$ & $ -0.0328$ \\
      $+4$ & $3$  & $191.1861$ & $ -0.1808$ & $ -0.4825$ & $ -0.5444$ & $ 0.3467$   &  $-0.4327 $ & $-0.0037$ & $ -0.2651$ & $  0.0252$ \\
      $+3$ & $2$  & $103.6268$ & $ -0.1383$ & $ -0.3375$ & $ -0.2391$ & $ 0.1617$   &  $-0.0780 $ & $ 0.0060$ & $ -0.0801$ & $  0.0282$ \\
      $+3$ & $4$  & $112.4585$ & $ -0.0894$ & $ -0.5304$ & $ -0.4781$ & $ 0.0894$   &  $-0.3264 $ & $-0.0021$ & $ -0.2007$ & $  0.0209$ \\
      $+2$ & $1$  & $ 51.3705$ & $ -0.0760$ & $ -0.1056$ & $ -0.0811$ & $ 0.0815$   &  $-0.0026 $ & $ 0.0200$ & $  0.0153$ & $  0.0541$ \\
      $+2$ & $3$  & $ 48.7002$ & $ -0.0627$ & $ -0.1445$ & $ -0.1185$ & $ 0.0349$   &  $-0.0297 $ & $-0.0035$ & $ -0.0223$ & $  0.0264$ \\
      $+2$ & $5$  & $ 56.1201$ & $ -0.0257$ & $ -0.5550$ & $ -0.4073$ & $-0.0454$   &  $-0.2398 $ & $ 0.0061$ & $ -0.1885$ & $ -0.0026$ \\
      $+1$ & $2$  & $ 17.1953$ & $ -0.0621$ & $ -0.0367$ & $ -0.0448$ & $-0.0097$   &  $-0.0183 $ & $-0.0206$ & $ -0.0102$ & $ -0.0005$ \\
  (IP)$+1$ & $4$  & $ 13.5653$ & $ -0.0142$ & $ -0.0438$ & $ -0.0436$ & $-0.0192$   &  $-0.0053 $ & $-0.0058$ & $ -0.0118$ & $  0.0083$ \\
      $+1$ & $6$  & $ 43.9706$ & $ -0.0290$ & $ -0.7264$ & $ -0.4557$ & $-0.1001$   &  $-0.2281 $ & $-0.0075$ & $ -0.2131$ & $ -0.0299$ \\
      $ 0$ & $1$  & $  2.1816$ & $ -0.0210$ & $  0.0057$ & $  0.0007$ & $ 0.0069$   &  $-0.0052 $ & $-0.0055$ & $  0.0039$ & $ -0.0017$ \\
      $ 0$ & $5$  & $  9.4860$ & $ -0.0182$ & $ -0.1259$ & $ -0.0750$ & $-0.0545$   &  $-0.0110 $ & $-0.0250$ & $ -0.0350$ & $ -0.0164$ \\
      $ 0$ & $7$  & $ 39.4276$ & $ -0.0247$ & $ -0.6879$ & $ -0.4390$ & $-0.0948$   &  $-0.2207 $ & $ 0.0025$ & $ -0.1962$ & $ -0.0152$ \\
 (-EA)$-1$ & $2$  & $ -1.4209$ & $  0.0017$ & $  0.0105$ & $  0.0092$ & $ 0.0259$   &  $-0.0083 $ & $ 0.0044$ & $ -0.0012$ & $ -0.0036$ \\
        \hline
	 MAD &   &             &   $0.0865$ & $  0.3275$ & $  0.2669$ & $0.1442$    &  0.1434      & $0.0078$  & $0.1303$   & $ 0.0192$ \\
	 \hline
	 \hline
    \end{tabular}
    \label{tab:oxygen}
\end{table*}

\begin{table}[h]
    \setlength{\tabcolsep}{12pt}
    \centering
    \caption{ECP parameters for Spectral Oxygen. The parametrization for each channel is given by $V_l(r) = \sum_k \beta_{lk} r^{n_{lk}-2}e^{-\alpha_{lk}r^2}$. The corresponding correlation consistent basis sets are included in the Supplementary Material. }
    \begin{tabular}{c r r r}
	\hline\hline
	Channel & $n_{lk}$ & $\alpha_{lk}$  & $\beta_{lk}$ \\
	\hline
	$p$   & 1  & 12.30997 &   6.00000 \\
	      & 3  & 14.76962 &  73.85984 \\
	      & 2  & 13.71419 & -47.87600 \\
	      \hline
	$s-p$ & 2  & 13.65512 & 85.86406 \\
	\hline\hline
    \end{tabular}
    \label{tab:pp_o}
\end{table}

\subsection{Sulfur}
\begin{figure}[ht!]
\caption{Potential energy surfaces of the S$_2$ molecule from CCSD(T). We have plotted the predictions from various treatments of the sulfur cores. Shown are the all-electron core (AE), all-electron uncorrelated core (UC), Burkatzki-Filippi-Dolg (BFD), Dirac-Fock Trail-Needs (TN) and the CCSD(T) spectrum matched  (Spectral) ECPs described in the text.}
\includegraphics[width=8cm]{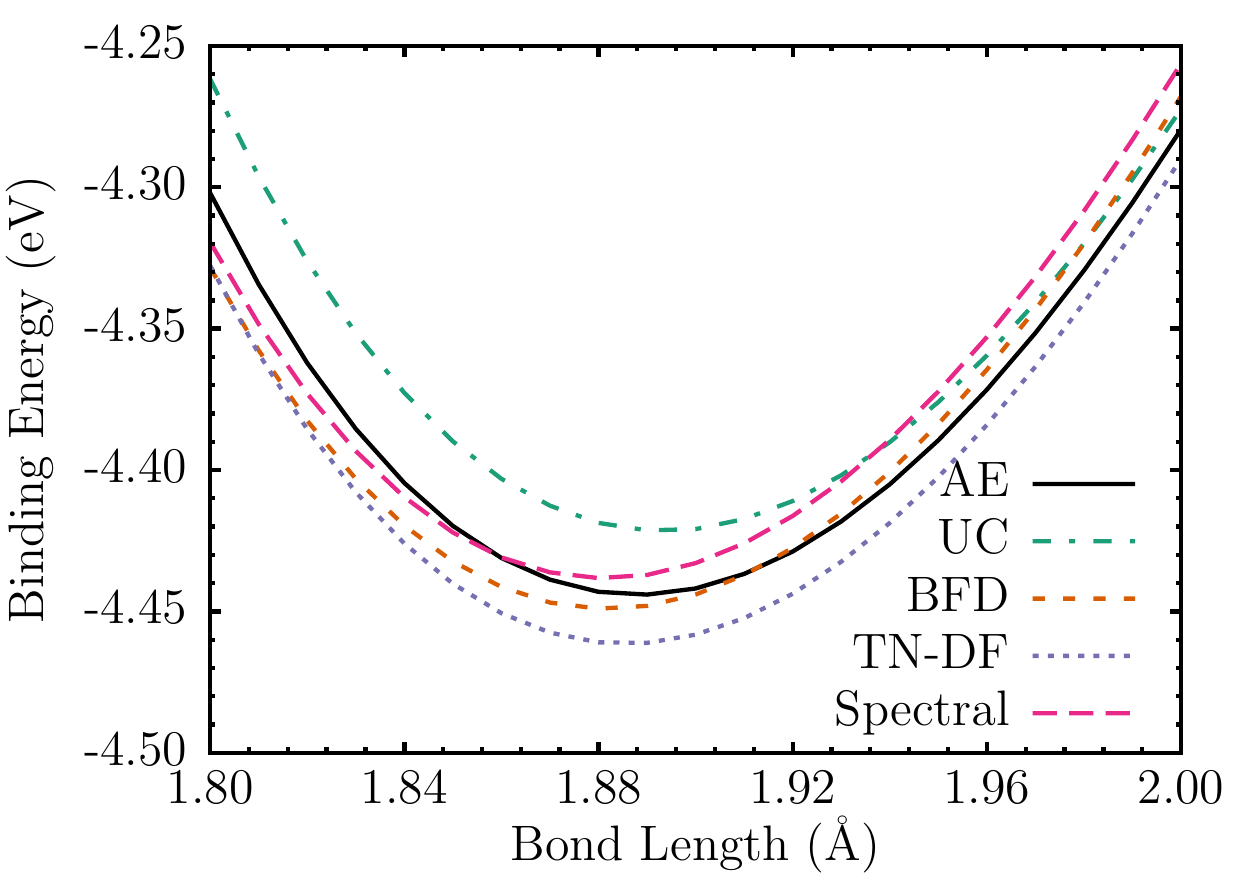}
\label{fig:s2_spectral}
\centering
\end{figure}

As a last illustration, we include our progress from the second-row atom, Sulfur.
Construction of the ECP for sulfur has followed similar steps as in previous cases. In Fig.\ref{fig:s2_spectral} we show the impact on the S$_2$ dimer its ground state ($^3\Sigma_g^{(-)}$)
of leaving the core uncorrelated on the overall accuracy and we see that near equilibrium 
its agreement with the all-electron prediction is at the level of $\approx0.03$~eV. 

\begin{table*}[ht!]
   \caption{Atomic data for Sulfur, similar to Table I. Energies in eV. Note that all gaps are calculated with reference to the ground state, namely Q=0 and 2S+1 = 3.}
   \begin{tabular}{rcrrrrrrrrrrrrrrrrrrrrrrr}
       \hline\hline
                       Q & 2S+1 &     AE Gap  &          UC  &         BFD  &        TN-DF  &  Spectral &  Spatial  & Spec/Space &  Constructed   \\
       \hline                                                                                                   
                    $+5$ &  $2$ & $188.3560$  &   $-0.4975$  &   $-0.3201$  & $-0.6243$  & $-0.0469$ & $-0.3620$ & $-0.9587$ & $-0.5141$ & & & &   \\
                    $+4$ &  $1$ & $115.7750$  &   $-0.2143$  &   $-0.0922$  & $-0.1412$  & $ 0.0123$ & $ 0.0693$ & $-0.4847$ & $-0.1838$ & & & &   \\
                    $+4$ &  $3$ & $126.1610$  &   $-0.3461$  &   $-0.1881$  & $-0.3452$  & $ 0.0580$ & $ 0.2173$ & $-0.3349$ & $-0.2871$ & & & &   \\
                    $+3$ &  $2$ & $ 68.5115$  &   $-0.1152$  &   $-0.0295$  & $-0.0381$  & $ 0.0290$ & $ 0.2974$ & $-0.1429$ & $-0.0926$ & & & &   \\
                    $+3$ &  $4$ & $ 77.3129$  &   $-0.2375$  &   $-0.1931$  & $-0.2359$  & $ 0.0318$ & $ 0.3573$ & $-0.0642$ & $-0.1816$ & & & &   \\
                    $+2$ &  $1$ & $ 35.1649$  &   $-0.1012$  &   $-0.0399$  & $-0.0338$  & $-0.0130$ & $ 0.2559$ & $-0.0326$ & $-0.0717$ & & & &   \\
                    $+2$ &  $3$ & $ 33.6900$  &   $-0.0515$  &   $-0.0332$  & $-0.0177$  & $-0.0076$ & $ 0.2565$ & $-0.0246$ & $-0.0582$ & & & &   \\
                    $+2$ &  $5$ & $ 40.8647$  &   $-0.1675$  &   $-0.2658$  & $-0.2325$  & $-0.0577$ & $ 0.2273$ & $-0.0197$ & $-0.1568$ & & & &   \\
                    $+1$ &  $2$ & $ 12.4609$  &   $-0.0920$  &   $-0.0732$  & $-0.0636$  & $-0.0673$ & $ 0.0794$ & $-0.0484$ & $-0.0830$ & & & &   \\
       (IP)$+1$ &  $4$ & $ 10.2976$  &   $-0.0159$  &   $-0.0581$  & $-0.0377$  & $-0.0536$ & $ 0.0826$ & $-0.0320$ & $-0.0559$ & & & &   \\
                    $+1$ &  $6$ & $ 30.7228$  &   $-0.1557$  &   $-0.2981$  & $-0.3143$  & $-0.0025$ & $ 0.2499$ & $ 0.1068$ & $-0.0134$ & & & &   \\
                    $+0$ &  $1$ & $  1.3319$  &   $-0.0399$  &   $-0.0029$  & $-0.0084$  & $-0.0015$ & $ 0.0090$ & $-0.0011$ & $-0.0094$ & & & &   \\
                    $+0$ &  $5$ & $  8.9838$  &   $-0.0223$  &   $-0.0563$  & $-0.0648$  & $-0.0244$ & $ 0.0749$ & $-0.0076$ & $-0.0136$ & & & &   \\
                    $+0$ &  $7$ & $ 27.8996$  &   $-0.1643$  &   $-0.3088$  & $-0.3504$  & $ 0.0094$ & $ 0.2134$ & $ 0.1109$ & $ 0.0170$ & & & &   \\
       (-EA)$-1$ &  $2$ & $ -2.0494$  &   $-0.0018$  &   $ 0.0277$  & $ 0.0064$  & $ 0.0302$ & $-0.0741$ & $-0.0124$ & $ 0.0149$ & & & &   \\
       \hline                                                                                                                           
       MAD               &      &             &   $ 0.1482$  &   $ 0.1325$  & $ 0.1676$  & $ 0.0297$ & $ 0.1884$ & $ 0.1588$ & $ 0.1169$ & & & &   \\
       \hline\hline 
   \end{tabular}
   \label{tab:sulfur}
\end{table*}

\begin{figure}[ht!]
\caption{Sulfur dimer binding energy discrepancies compared to the all-electron CCSD(T) binding curve.} 
\includegraphics[width=8cm]{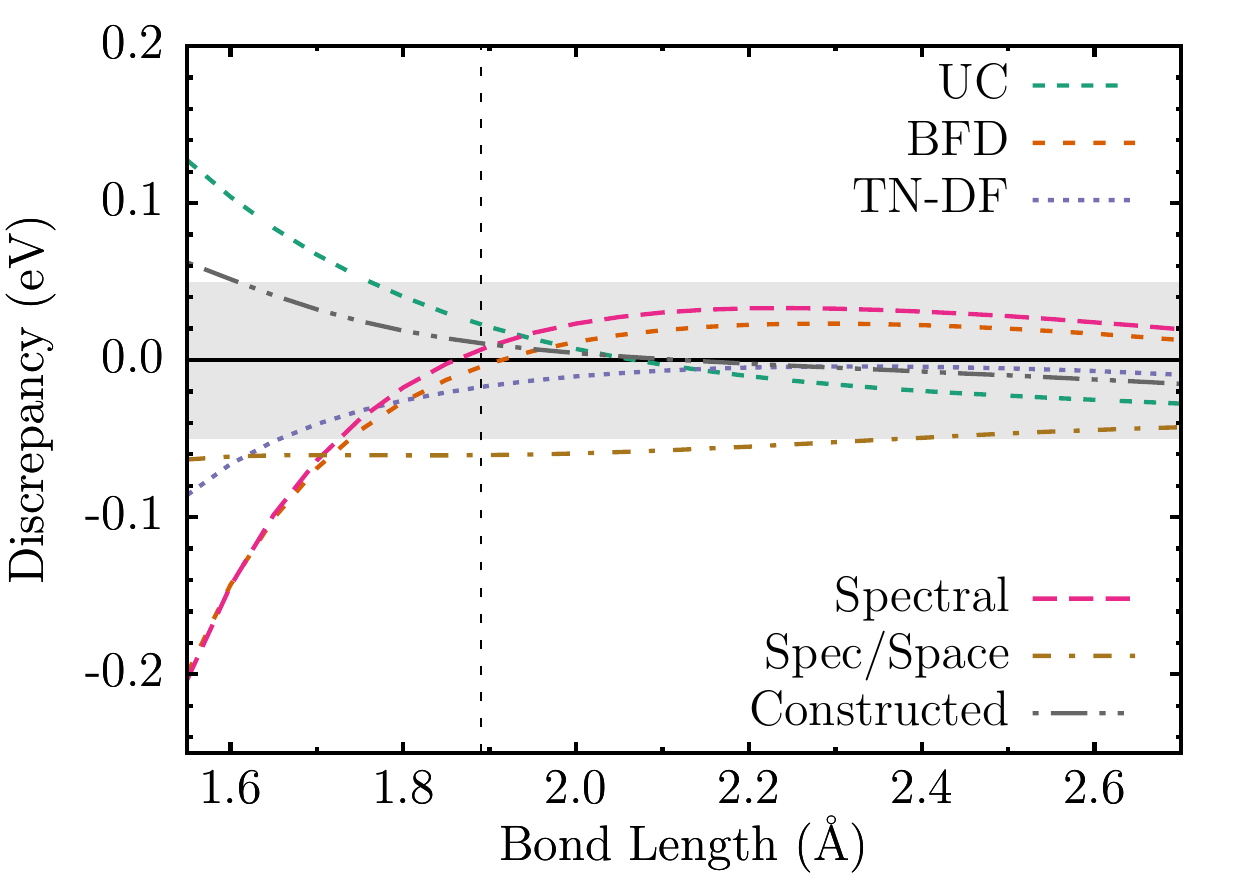}
\label{fig:s2_all_diffs}
\centering
\end{figure}

\begin{figure}[ht!]
\caption{Sulfur dimer binding energy discrepancies compared to the all-electron CCSD(T) binding curve.}
\includegraphics[width=8cm]{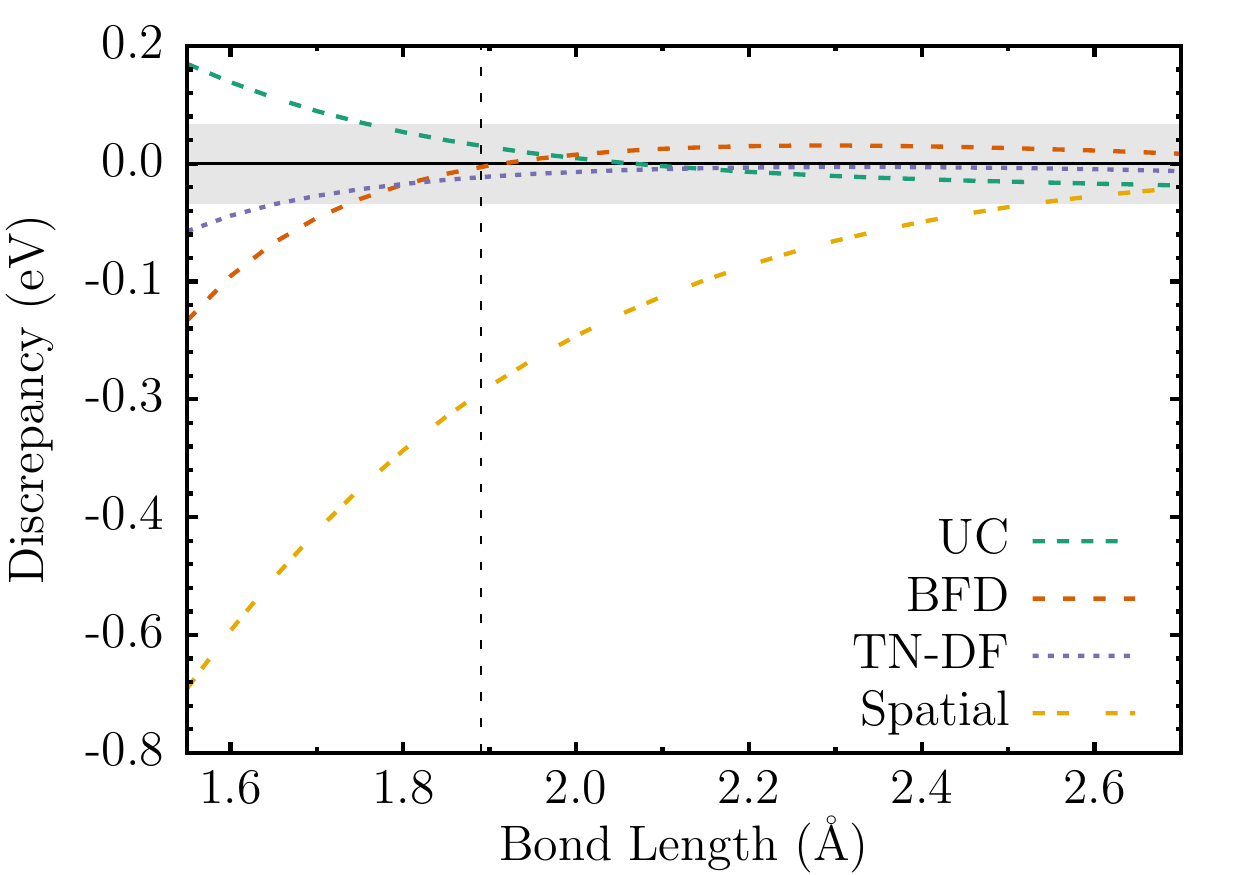}
\label{fig:s2_spatial_diffs}
\centering
\end{figure}

For this case, we utilized a Ne core ECP and attempted a set of optimization strategies similar to the investigation of N and O.
In this case, we again consider constructing ECPs to mimic all-electron many-body spectral, spatial and combined properties, in turn.
Figures \ref{fig:s2_all_diffs} and \ref{fig:s2_spatial_diffs} show CCSD(T) binding energy discrepancies from the all-electron S$_2$ molecule ($^3\Sigma_g^{(-)}$) for various approximations to the sulfur cores including our generated ECPs.

For the spectral case, we again have used an all-electron reference formed from bound excitations and included all possible valence ground states of total spin and charge whereby we minimized the spectral discrepancies exclusively with respect to ECP's parameters.
In Table \ref{tab:sulfur}, it is shown that the MAD from the all-electron excitation energies that is no more than $0.03$~eV in this case. 
This agreement is nearly an order of magnitude improvement over the other approximations to the core.
Additionally, with the spectral objective function, we observe that the agreement with the all-electron binding energy at equilibrium is very good with an error of no more than $0.01$~eV.
For shorter separations of the dimer, however, the spectral ECP undershoots the all-electron binding energy by tenths of eV and as a result of this large change from equilibrium, we see a non-negligible slope and curvature in the discrepancy at equilibrium which negatively impacts the agreement with the all-electron prediction of the ground state vibrational frequency.

Using the spatial information for the same set of states, we generated the single-body density matrices from all-electron CISD wave functions and subsequently imposed that the single-body density matrix of the pseudoatom's CISD wave function match beyond a core radius.
The resulting ECPs constructed from this procedure were not adequately transferable which we attribute again to marginal electronic correlations signal in the natural orbitals as we argued for the cases of N and O.
It can also be observed from Table \ref{tab:sulfur} that the atomic properties are generally negatively affected by optimizing spatial data alone.

For sulfur, we also attempted to match both the all-electron excitation energies and single-body density matrices, simultaneously.   
For this case, the binding energy discrepancies from the all-electron atom are shown in Fig. \ref{fig:s2_all_diffs}.
Here we see that the error is quite uniform over the entire region plotted, and moreover, it is no more than about $0.06$ eV.

Lastly, we again considered additional molecular constraints on the objective function in order to further improve the ECP's dimer properties as described for the nitrogen atom. 
The constraints were constructed in such a way as to reach a balance between the spectral and molecular properties. 
We show its discrepancies in Fig. \ref{fig:s2_all_diffs}. 
For this ECP, we see that the dimer's error is mostly within $0.05$ eV throughout the plotted region and the spectral properties are an improvement over both UC and previously generated ECPs.
The parameters for sulfur's constructed ECP are shared in Table \ref{tab:pp_s}. 

\begin{table}[h]
    \setlength{\tabcolsep}{12pt}
    \centering
    \caption{ECP parameters for Constructed ECP for Sulfur. The parametrization for each channel is given by $V_l(r) = \sum_k \beta_{lk} r^{n_{lk}-2}e^{-\alpha_{lk}r^2}$. The corresponding correlation consistent basis sets are included in the Supplementary Material. }
    \begin{tabular}{c r r r}
	\hline\hline
	Channel & $n_{lk}$ & $\alpha_{lk}$  & $\beta_{lk}$ \\
	\hline
	$d$  & 1 &  4.23812  &   3.06000  \\
         & 1 &  2.19773  &   2.94000  \\
         & 3 &  1.71348  &  12.96866  \\
         & 3 & 10.20072  &   6.46132  \\
         & 2 &  3.41487  & -10.45671  \\
         & 2 &  1.40439  &  -9.79751  \\
    \hline
    $s-d$  & 2 & 3.91958 &  23.19840 \\
           & 2 & 3.91388 &  22.28866 \\
    \hline       
    $p-d$  & 2 & 2.71232 &   8.39601 \\
           & 2 & 3.20078 &  11.15610 \\    
	\hline\hline
    \end{tabular}
    \label{tab:pp_s}
\end{table}

\section{Transferability}\label{transferability}
One of the desired properties for our ECPs is transferability, i.e., high-quality performance in systems that were not used in our optimization procedure.
In the case of our spectral ECPs, the optimization of the parameters only involved atomic properties; for these ECPs, the dimer discrepancies compared to the all-electron results illustrate the transferability of our ECPs to a degree. 
However, for our constructed ECPs, the parameters were tuned in order to produce improved dimer properties while not sacrificing the atomic properties and thus the transferability should be verified also on independent examples. 
Therefore, it is desirable to test those cases in other bonding environments in order to illustrate their transferability beyond the dimers. 
For this purpose, we have calculated potential energy surface discrepancies for hydrides,  oxides and a handful of additional molecules BH$_3$, BN, BS and CN and in order to ascertain their level of transferability.
All molecular calculations were performed with either the \textsc{Molpro} quantum chemistry package \cite{MOLPRO-WIREs} or the \textsc{Gaussian09} code \cite{g09}.

\begin{figure}[ht!]
\caption{NH binding energy discrepancies for various ECPs}
\includegraphics[width=8cm]{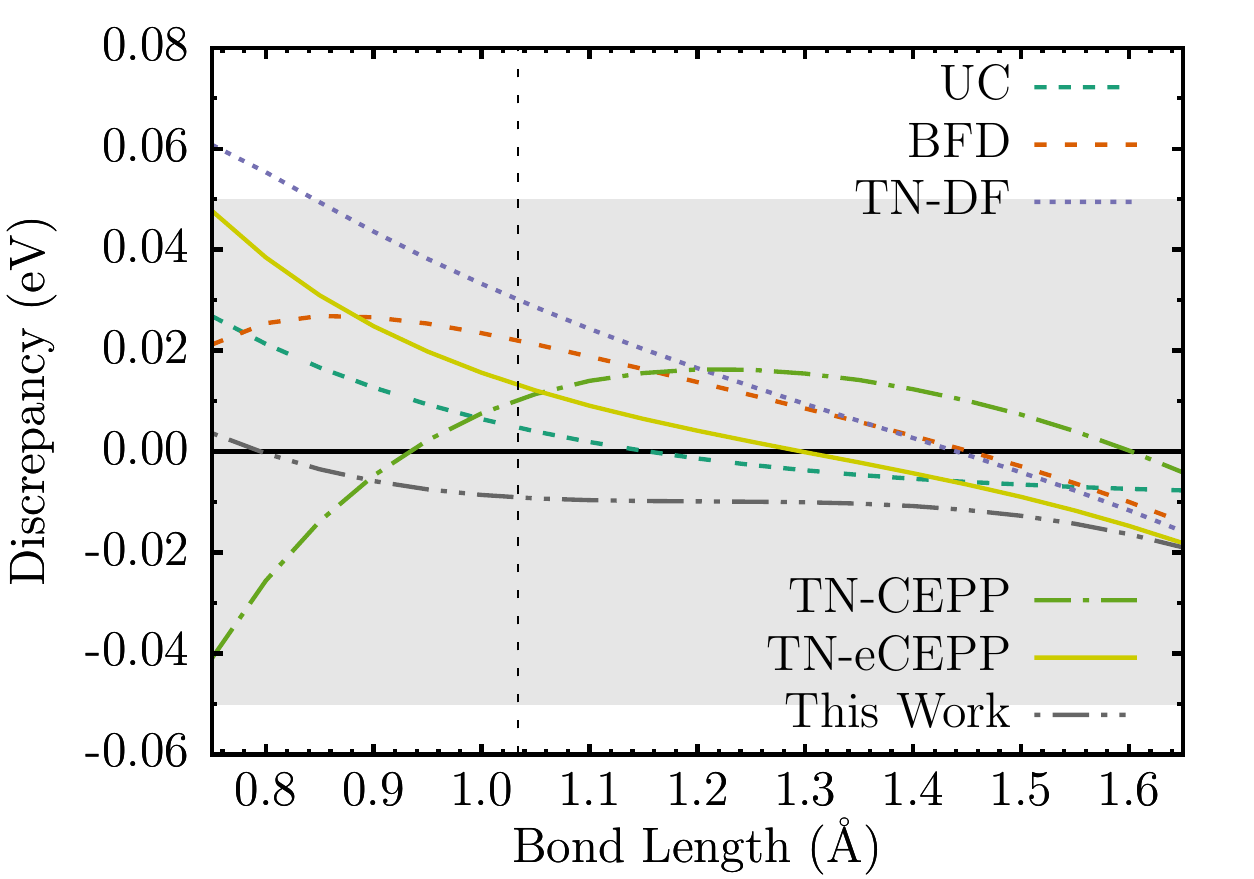}
\label{fig:nh_diffs}
\centering
\end{figure}

\begin{figure}[ht!]
\caption{OH binding energy discrepancies for various ECPs. For Oxygen, we use our spectral ECP.}
\includegraphics[width=8cm]{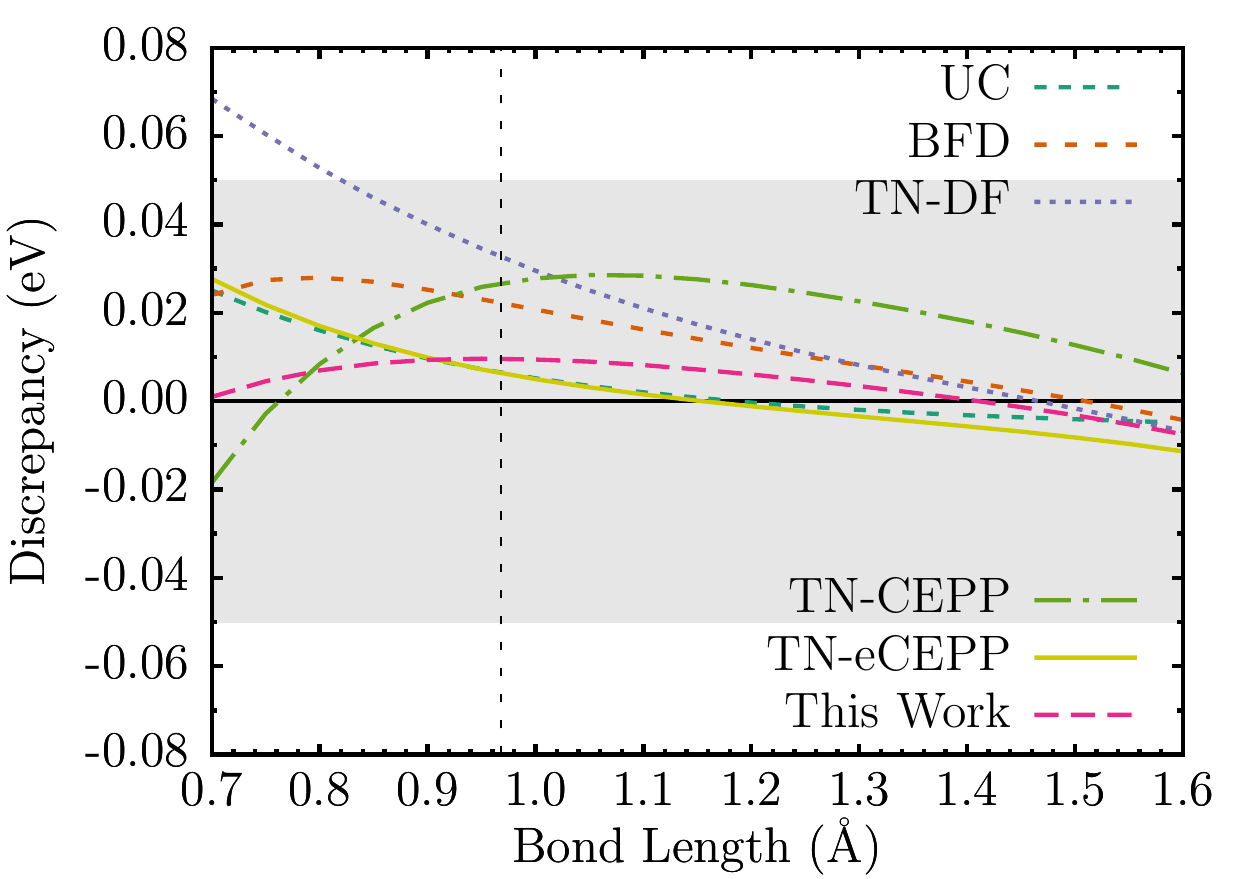}
\label{fig:oh_diffs}
\centering
\end{figure}

\begin{figure}[ht!]
\caption{NO binding energy discrepancies for various ECPs. For Nitrogen, we use our constructed ECP and for Oxygen, we use our spectral ECP.}
\includegraphics[width=8cm]{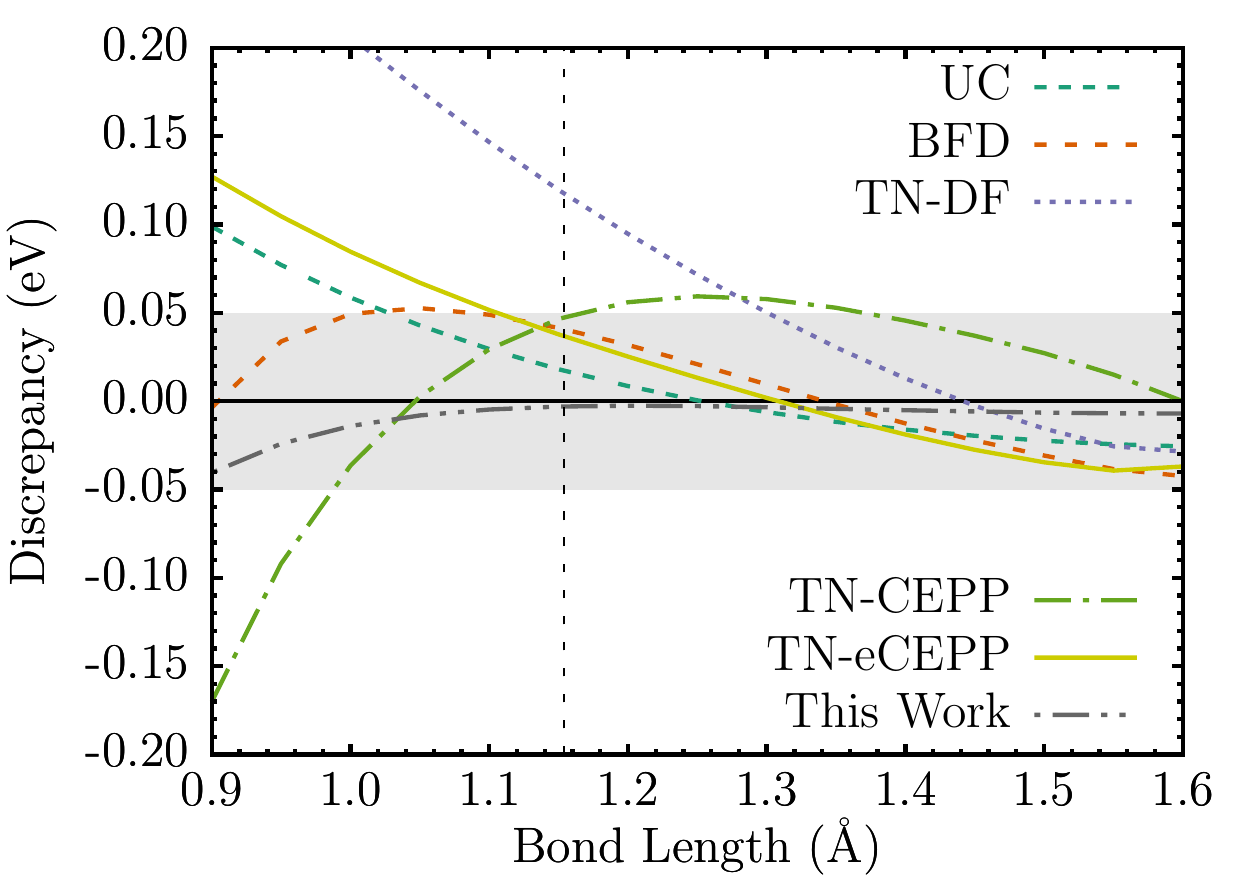}
\label{fig:no_diffs}
\centering
\end{figure}

\begin{figure}[ht!]
\caption{SH binding energy discrepancies compared to the all-electron CCSD(T) binding curve. For Sulfur, we use our constructed ECP and for Oxygen, we use our spectral ECP.}
\includegraphics[width=8cm]{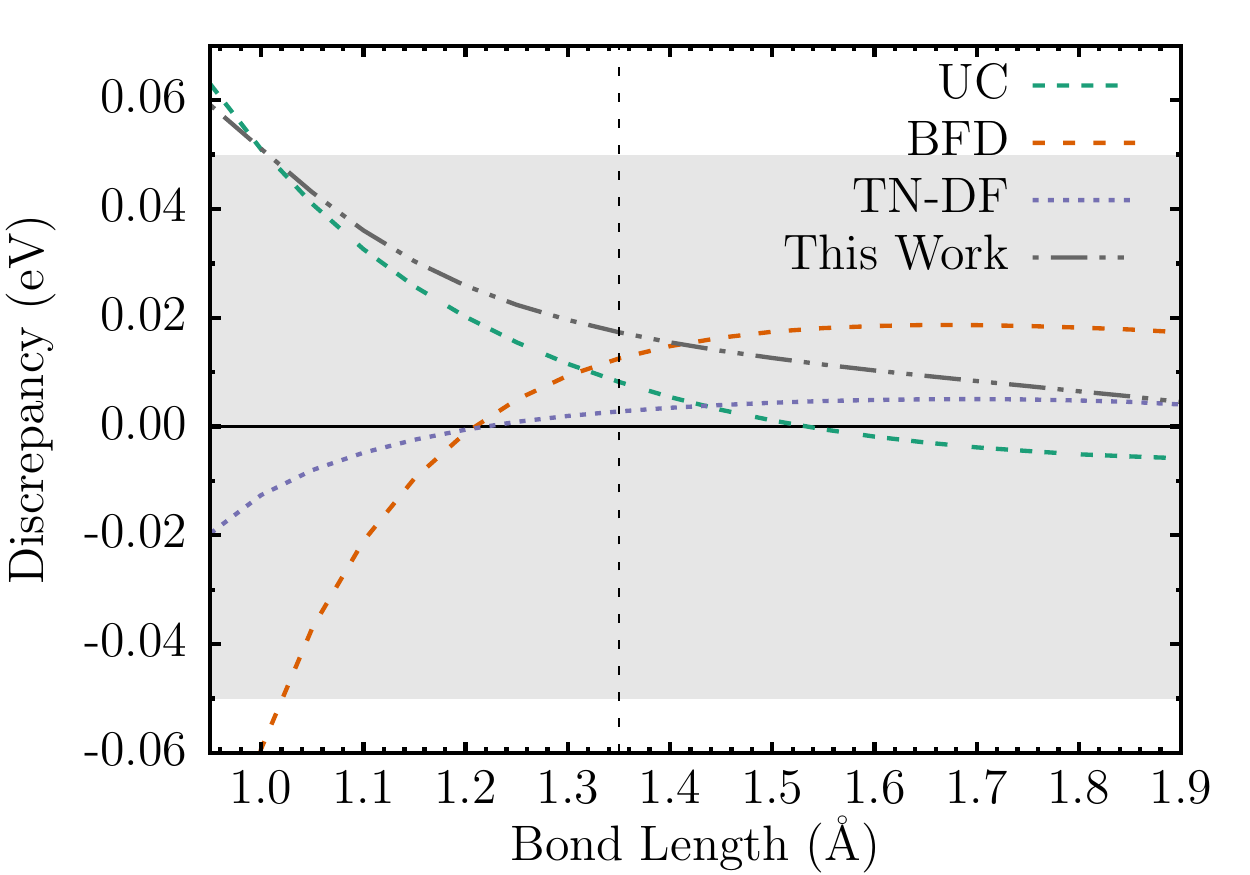}
\label{fig:sh_diffs}
\centering
\end{figure}

\begin{figure}[ht!]
\caption{SO binding energy discrepancies compared to the all-electron CCSD(T) binding curve. For Sulfur, we use our constructed ECP.}
\includegraphics[width=8cm]{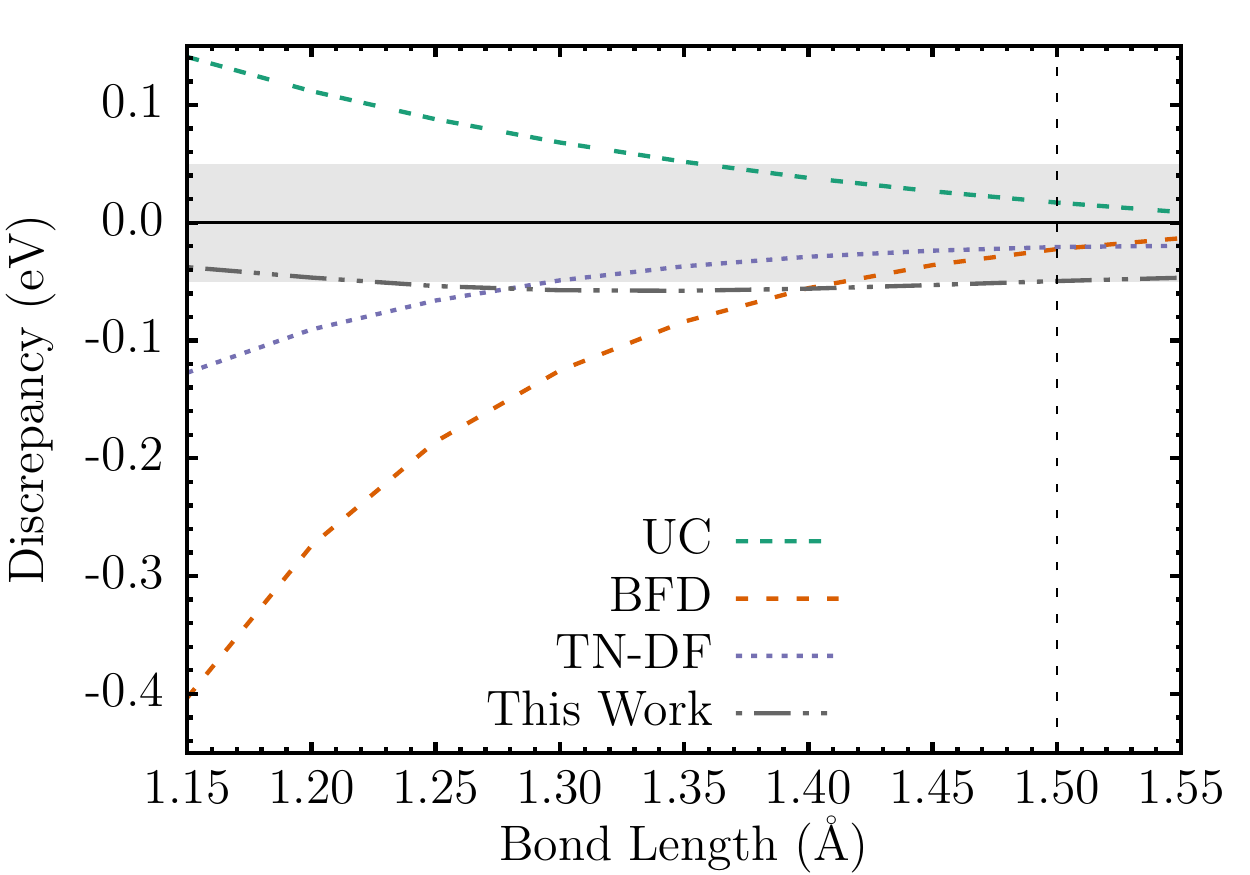}
\label{fig:so_diffs}
\centering
\end{figure}

\begin{figure*}[ht!]
\centering
\caption{Discrepancies of molecular binding parameters of our ECPs, UC and previous constructions with respect to all-electron CCSD(T) calculations. Parameters were obtained from Morse potential fits in all cases. 
}
	    \begin{subfigure}[b]{0.5\textwidth}
		\centering
		\includegraphics[width=\linewidth]{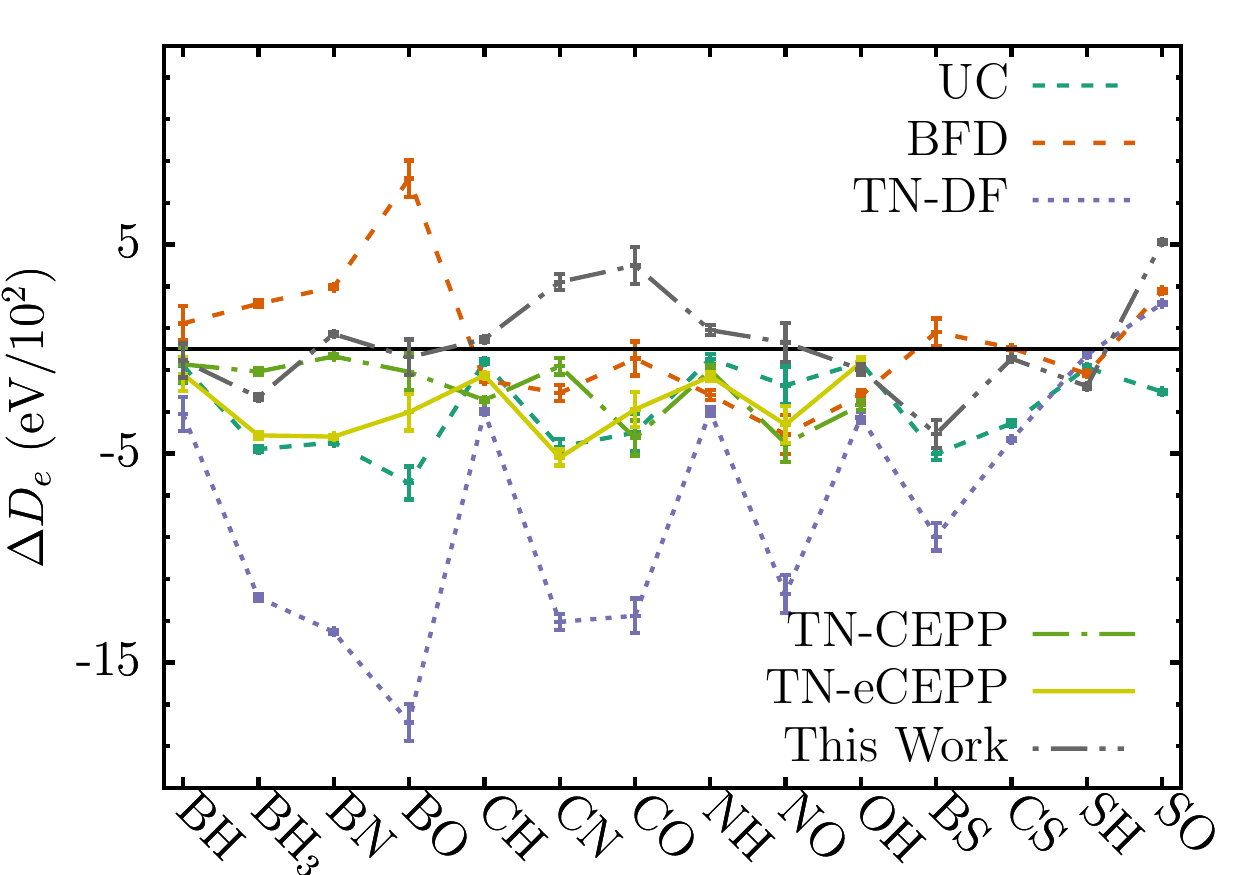}
		\caption{Dissociation energy discrepancies}
	    \end{subfigure}
	    \begin{subfigure}[b]{0.5\textwidth}
		\centering
		\includegraphics[width=\linewidth]{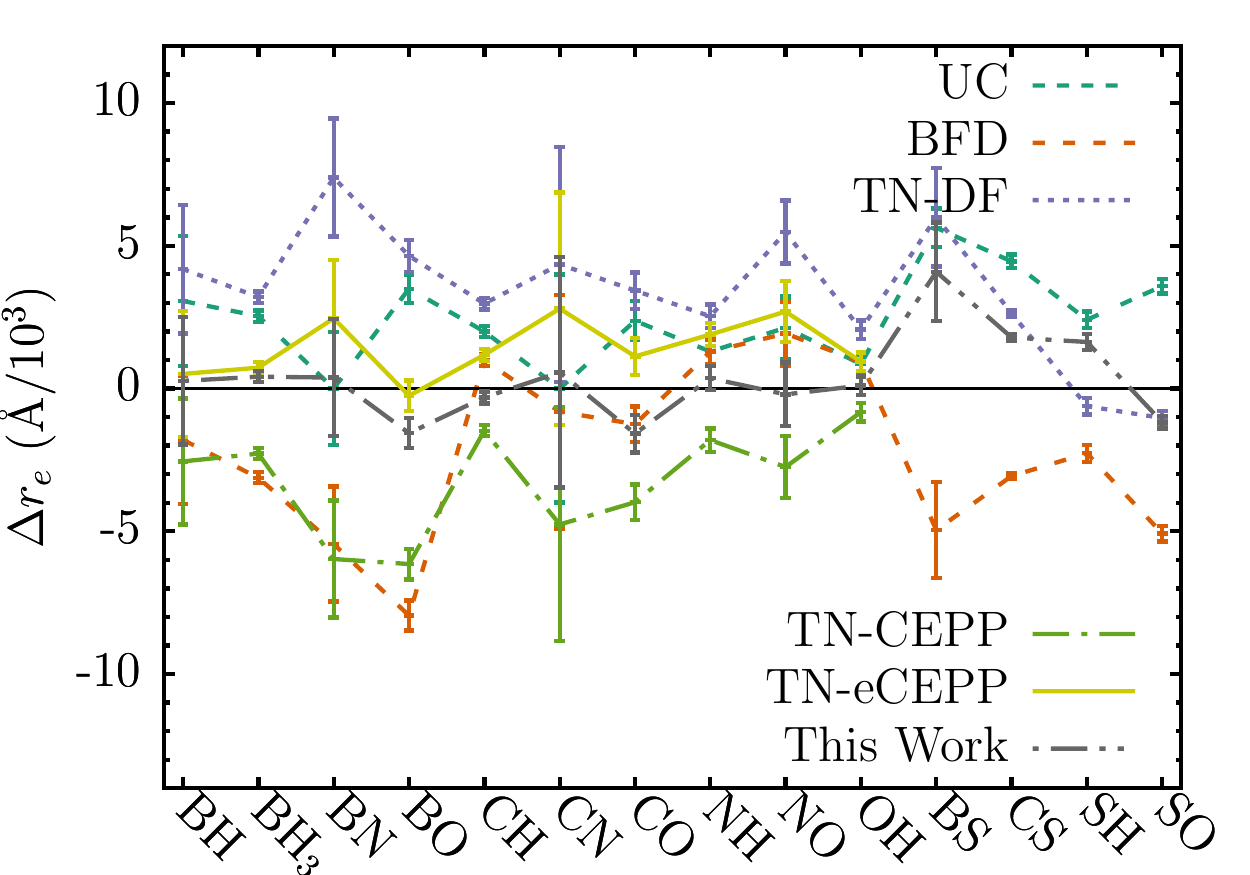}
		\caption{Equilibrium bond length discrepancies.}
	    \end{subfigure}
	    \begin{subfigure}[b]{0.5\textwidth}
		\centering
		\includegraphics[width=\linewidth]{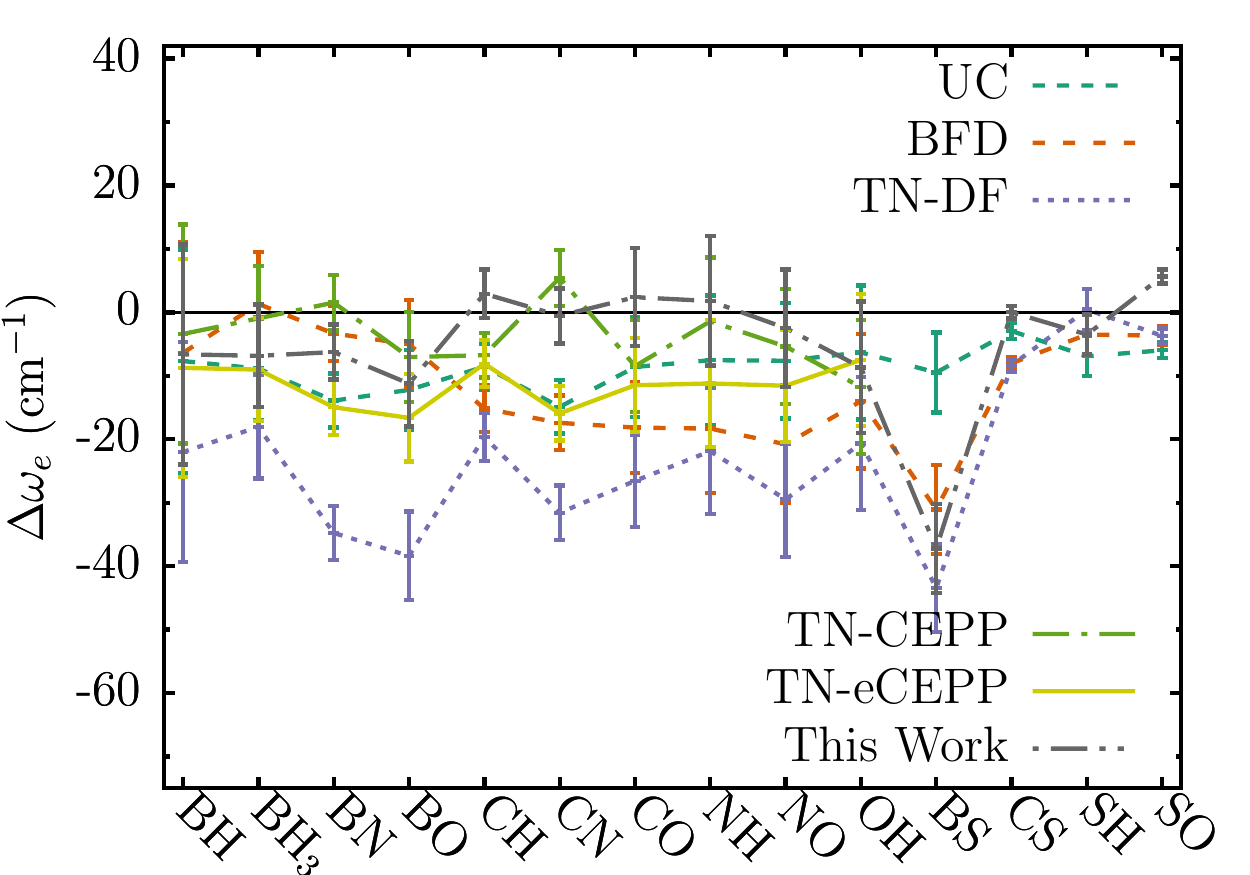}
		\caption{Vibrational frequency discrepancies.}
	    \end{subfigure}
\label{fig:global_transferability}
\end{figure*}

We show binding energy discrepancies with respect to all-electron CCSD(T) for the NH, OH, NO, SH and SO molecules in Figs. \ref{fig:nh_diffs},\ref{fig:oh_diffs},\ref{fig:no_diffs},\ref{fig:sh_diffs} and \ref{fig:so_diffs}, respectively. 
Furthermore, we summarize the discrepancies in the binding parameters, $D_e$, $r_e$, and $\omega_e$ of all molecules considered in this work with respect to all-electron CCSD(T) values in Fig. \ref{fig:global_transferability}, where $D_e$ is the dissociation energy, $r_e$ is the equilibrium bond length, and $\omega_e$ is the vibrational frequency.
For each case, the parameters and their errors were obtained from fitting the potential energy surface to the Morse potential near equilibrium, where the potential is given as
\begin{equation}
    V(r) = D_e(e^{-2 a (r-r_e)}-2e^{-a(r-r_e)})
    \label{eqn:morse}
\end{equation}
where $a$ is related to the vibrational frequency by
\begin{equation}
    \omega_e = \sqrt{\frac{2 a^2 D_e}{\mu}}
    \label{eqn:freq}
\end{equation}
and $\mu$ is the reduced mass of the molecule.

To make the comparison of the pseudoatom transferabilities easier, we also share the MADs of these parameters, both at equilibrium and at the dissociation limit at short bond lengths corresponding to high pressures, in Table \ref{tab:global_mads}.
From the table, it is shown that our ECPs perform the best overall, where the mean-absolute deviation from the all-electron dissociation energy is smaller than all considered core approximations;
furthermore, this improvement 
and better overall balance is achieved with 
very limited variational freedom as given by
the choice of the ECP form. 
The only exception is a marginally better MAD for the vibrational frequencies of TN-CEPP with 5(3) versus ours 7(2) cm$^{-1}$.

\begin{table*}[ht!]
    \centering
    \caption{Mean absolute deviations of discrepancies of binding parameters at equilibrium ($D_e$, $r_e$ and $\omega_e$) and near the dissociation threshold ($D_{diss}$) at short bond lengths for our ECPs and previous constructions with respect to all-electron CCSD(T) calculations. The system sets correspond to Fig.\ref{fig:global_transferability} except for the BH$_3$ molecule which was omitted from the MADs of the dissociation threshold energy.}
    \begin{tabular}{c | cccccc}
	\hline
	\hline
	                          &          UC &        BFD &      TN-DF &    TN-CEPP &   TN-eCEPP &  This Work \\
	\hline
	 $D_e$      (eV/$10^2$)     &    2.9(1) &     2.3(1) &     7.8(1) &     1.9(2) &     2.7(2) &   1.8(1)   \\
	 $r_e$      (\AA/$10^3$)    &    2.9(2) &     2.9(2) &     3.6(2) &     3.2(3) &     1.1(2) &   1.0(2)   \\
	 $\omega_e$ (cm$^{-1}$)     &      9(2) &      12(2) &      23(2) &       5(3) &      12(3) &     7(2)   \\
	 $D_{diss}$      (eV/$10^2$)  &  11.94  &      23.76 &      9.78  &       22.64  &  7.10              &     6.13\\
	 \hline
	 \hline
    \end{tabular}
    \label{tab:global_mads}
\end{table*}

\section{Conclusions}\label{conclusions}
 
Our paper presents an advancement in the construction of effective core potentials for accurate, correlated valence-only calculations. 
A key difference from previous constructions being the consistent use of nearly exact many-body approaches to build the ECPs
and balancing this 
with refinement by molecular 
data that improves the transferability.
We introduce isospectrality of all-electron and ECP Hamiltonians on the subspace of valence states as a foundation in formulating the objective functions. 
Additional criteria that were explored included many-body spatial information from density matrices. 
This was followed by analyzing the accuracy on dimers and if significant differences were observed we included the dimer information to boost the accuracy and transferability.
This was done in an iterative manner and combined constructions were used for efficiency reasons. 
We were especially careful in obtaining consistent binding curves within a large portion of the bond, i.e., from separations near the dissociation limit to at least equilibrium.
We further ascertained the transferability in those cases on hydride, oxide molecules and selected other molecules that involved the considered element.

The calculations are done at the relativistic level, therefore, energy differences are very close to the actual experimental values for the relevant quantities. 

We were able to decrease the atomic spectral errors for B, C, N, O, and S atoms to significantly smaller values when compared to previous constructions.

Interestingly, the spectral discrepancies are also {\it smaller} than in uncorrelated core calculations. 
Therefore the constructed operators are able to effectively accommodate at least some of the effects from core-core and core-valence correlations as it is clear from comparisons with uncorrelated core calculations.
Similar argument applies to the improved description of simple molecular systems for a range of bond lengths including very short bonds that are relevant for calculations of materials at high pressures. 
Rather surprisingly, this implies that for a range of valence properties these ECPs will provide the same or {\it even more accurate} results then all-electron treatments with uncorrelated cores. 
Note the caveat ``for a range of valence properties" in the previous statement since this is true only for states/properties that are not too far from the one used in optimization.
Clearly, inclusion of larger range of energies and states that can be described would require either additional terms in the effective ECP Hamiltonian or, in general, including more subshells into the valence space (as is common for heavier elements).

Our experience from these constructions shows:

a) the accuracy of even the simplest forms of semi-local ECPs can be boosted very significantly by appropriate many-body constructions;

b) optimization is quite involved since minor changes in the inputs and the objective function (e.g., increasing the size of the basis set, boosting the spectral accuracy from CCSD to CCSD(T), etc) 
complicate the optimizations with many local minima solutions, i.e., we see a clear hallmark of an ill-conditioned problem; 

c) there are differences from element to element influenced by the extent and choice of excitations, occupations and other details showing thus further  complications from the ill-conditioned character of the task.

The obtained results and experience offers a route for constructing comparable quality
ECPs for more involved elements such as $3d$ transition series and beyond. Some adjustments in the constructions and expanding the parameter sets can be expected. Overall prospects for heavier elements would be limited basically only by the availability of 
codes that can do accurate correlated all-electron relativistic calculations for atomic and small molecular systems. 

In this work we have pursued the construction of ECPs from a many-body framework as opposed to one-particle schemes.
In addition, we have included further considerations into the
construction such as molecular systems at and away of the equilibrium for better overall transferability.

One-particle constructions have generally involved norm/shape and charge conservations, one-particle eigenvalues, logarithmic derivatives, differences between approximate total energies, etc.
Unfortunately, all of these were subject to the biases of the (approximate) method used to solve the underlying atomic problem. 
The expectation and hope were that in subsequent calculations the systematic cancellations of the underlying biases would be the same as in all-electron setting. 
In many cases, this worked reasonably well. However, this was not true systematically and for some elements, states and bonding environments the errors are significant. 
For example, in DFT all-electron and ECP calculations for just a single atom could lead to large differences. In addition,
special adjustments might be necessary such as the nonlinear core corrections for transition metal atoms \cite{Louie:1982prb}, etc. 
These complications, however, defy the universal use of such ECPs in other methods, what is one of our stated goals. 
In addition, there has always been a lingering (and valid) question why an ECP constructed in DFT or HF methods should be any good in a correlated many-body approach. 
Clearly, that was one of our motivations to use a many-body framework consistently throughout the construction, in testing and also in further developments. 

The validity and limits of the constructed ECPs also become much more transparent if the {\it information about systematic errors built into ECPs is provided upfront} so as to clear the stage for subsequent calculations. 
The simplicity of the basic ECP form is also highly desirable since that enables wide use in many settings and approaches. 

Note that presented ECPs are of the simplest semilocal type with minimal form and size, with only one or two gaussians per channel. 
It is realistic to expect that more careful and more elaborate fits could further tune the properties beyond accuracies we have currently obtained. 
The corresponding optimizations are still challenging and are hampered by complicated couplings between the atomic solvers and optimization methods and require further refinement, higher efficiency, and more robustness.  
At the same time, the presented results offer encouraging examples of the accuracy that can be obtained with present day capabilities and show that there is significant room for further improvements and expansions.

Input and output files for the calculations performed in this study are available \cite{data_doi} via the Materials Data Facility.

\section*{Supplementary Material}
See supplementary material for basis set extrapolation analysis and a listing of correlation consistent basis sets for each pseudoatom presented in this work.

\smallskip

{\bf Acknowledgments.}

We are grateful for stimulating discussions with Jerry Whitten during this study. We also would like to thank Paul Kent for a careful reading of the manuscript while in preparation which lead to a number of vital improvements.

The majority of this work (development of the methods, calculations, tests, and writing of the paper) has been supported by 
 the U.S. Department of Energy, Office of Science, Basic Energy Sciences, Materials Sciences and Engineering Division, as part of the Computational Materials Sciences Program and Center for Predictive Simulation of Functional Materials.
The initial theoretical and conceptual considerations were supported by ORNL/UT Batelle, LLC, subcontract N. 4000144475.

The calculations for this work were performed mostly at Sandia National Laboratories, while some of the calculation have been carried out at TACC under XSEDE.

Sandia National Laboratories is a multimission laboratory managed and operated by National Technology and Engineering Solutions of Sandia LLC, a wholly owned subsidiary of Honeywell International Inc. for the U.S. Department of Energy National Nuclear Security Administration under contract DE-NA0003525.

\bibliography{main.bib}

\begin{thebibliography}{37}
\bibitem[{foo()}]{foot1}
\bibinfo{note}{In this work, we neglect this distinction and use the terms ECP
  and pseudopotential interchangeably.}
\end{thebibliography}
\end{document}